\documentclass[fleqn,usenatbib]{mnras}

\usepackage{newtxtext,newtxmath}

\usepackage[T1]{fontenc}

\DeclareRobustCommand{\VAN}[3]{#2}
\let\VANthebibliography\thebibliography
\def\thebibliography{\DeclareRobustCommand{\VAN}[3]{##3}\VANthebibliography}

\usepackage{graphicx, float, hyperref, tabularx, physics, ulem}	
\usepackage{amsmath}	
\usepackage{amssymb}	

\usepackage{threeparttable} 

\newcommand{\grs}{{GRS~1915+105 }}
\newcommand{\maxi}{{MAXI~J1535$-$571 }}
\newcommand{\rxte}{{\it{RXTE} }}
\newcommand{\hxmt}{{\textit{Insight}-HXMT} }

\title[Comptonization region of \maxi]{The evolution of the corona in \maxi through type-C quasi-periodic oscillations with \textit{Insight}-HXMT}

\author[Y.\ Zhang et al.]{
Yuexin Zhang,$^{1,2}$\thanks{E-mail: yzhang@astro.rug.nl}
Mariano M\'{e}ndez,$^{1}$
Federico Garc\'{i}a,$^{1,8}$
Shuang-Nan Zhang,$^{2,5}$
Konstantinos Karpouzas,$^{1,3}$
\newauthor
Diego Altamirano,$^{3}$
Tomaso M.\ Belloni,$^{4}$
Jinlu Qu,$^{2,5}$
Shu Zhang,$^2$
Lian Tao,$^2$
Liang Zhang,$^{3,2}$
Yue Huang,$^{2}$
\newauthor
Lingda Kong,$^{2,5}$
Ruican Ma,$^{2,5}$
Wei Yu,$^{2,5}$
Divya Rawat,$^{6,7}$
and Candela Bellavita$^{9}$
\\
$^{1}$Kapteyn Astronomical Institute, University of Groningen, P.O. BOX 800, 9700 AV Groningen, The Netherlands\\
$^{2}$Key Laboratory of Particle Astrophysics, Institute of High Energy Physics, Chinese Academy of Sciences, Beijing 100049, People's Republic of China\\
$^{3}$School of Physics and Astronomy, University of Southampton, Southampton, SO17 1BJ, UK\\
$^{4}$INAF-Osservatorio Astronomico di Brera, via E.\ Bianchi 46, I-23807 Merate, Italy\\
$^{5}$University of Chinese Academy of Sciences, Chinese Academy of Sciences, Beijing 100049, People's Republic of China\\
$^{6}$Department of physics, IIT Kanpur, Kanpur, Uttar Pradesh 208016, India\\
$^{7}$Inter-University Center for Astronomy and Astrophysics, Ganeshkhind, Pune 411007, India\\
$^{8}$Instituto Argentino de Radioastronom\'ia (CCT La Plata, CONICET; CICPBA; UNLP), C.C.5, (1894) Villa Elisa, Buenos Aires, Argentina\\
$^{9}$Facultad de Ciencias Astron\'omicas y Geof\'{\i}sicas, Universidad Nacional de La Plata, Paseo del Bosque, B1900FWA La Plata, Argentina\\
}

\date{Accepted XXX. Received YYY; in original form ZZZ}

\begin{document}
\label{firstpage}
\pagerange{\pageref{firstpage}--\pageref{lastpage}}
\maketitle

\begin{abstract}
    Type-C quasi-periodic oscillations (QPOs) in black hole X-ray transients can appear when the source is in the low-hard and hard-intermediate states. The spectral-timing evolution of the type-C QPO in \maxi has been recently studied with \textit{Insight}-HXMT. Here we fit simultaneously the time-averaged energy spectrum, using a relativistic reflection model, and the fractional rms and phase-lag spectra of the type-C QPOs, using a recently developed time-dependent Comptonization model when the source was in the intermediate state. We show, for the first time, that the time-dependent Comptonization model can successfully explain the X-ray data up to 100~keV. We find that in the hard-intermediate state the frequency of the type-C QPO decreases from 2.6~Hz to 2.1~Hz, then increases to 3.3~Hz, and finally increases to $\sim$ 9~Hz. Simultaneously with this, the evolution of corona size and the feedback fraction (the fraction of photons up-scattered in the corona that return to the disc) indicates the change of the morphology of the corona. Comparing with contemporaneous radio observations, this evolution suggests a possible connection between the corona and the jet when the system is in the hard-intermediate state and about to transit into the soft-intermediate state.
\end{abstract}




\section{Introduction}\label{sec:intro}

Black hole low-mass X-ray binaries (LMXBs) consist of a low-mass star and a central black hole which accretes materials from the star via Roche-lobe overflow. The inner part of an accreting black hole system includes an accretion disc and a hot Comptonization region, which is called the corona~\citep[for a review, see][]{2010LNP...794...17G}. As the accretion disc is close to the black hole, the infalling gas releases strong gravitational energy, producing a multi-temperature blackbody emission that can be detected with a temperature around 0.3--2.0 keV in the soft X-ray band. Some of the soft photons are inverse Comptonized into hard X-ray photons in the corona, producing roughly a power-law continuum in the energy spectrum~\citep[see][for a review]{2007A&ARv..15....1D}. A fraction of the Comptonized photons illuminate the inner accretion disc and can be Compton back-scattered in the disc, producing a relativistic reflection spectrum with characteristic emission lines, among which the most prominent feature is a broad iron $K_{\alpha}$ line around 6.4~keV, and a Compton hump at around 20~keV~\citep{1989MNRAS.238..729F,2014ApJ...782...76G}. Due to the relativistic property of the spacetime, the X-ray reflection spectrum has recently been developed as a powerful tool to test General Relativity~\citep[for a review, see][]{2021SSRv..217...65B}.

A typical outburst of a black hole X-ray transient follows a `q' path in the hardness intensity diagram~\citep[HID;][]{2001ApJS..132..377H,2004MNRAS.355.1105F}, showing a hysteresis phenomenon between the soft and hard states~\citep{2003MNRAS.338..189M}. Before the outburst, the source spends time in quiescence with the X-ray luminosity being several orders of magnitude lower than during outburst. As the outburst starts, the source enters the low-hard state with a dominant hard-photon emission; as mass accretion rate from the disc onto the black hole increases the source at some point quickly (on the order of days) transits to the high-soft state, during which the X-ray spectrum is disc dominated and the power-law emission becomes steeper; finally the source returns back to the low-hard state and eventually to quiescence as mass accretion rate decreases; a relativistic jet can appear in the transitions between the hard and the soft states~\citep{2004MNRAS.355.1105F,2006ARA&A..44...49R}. Usually two types of relativistic jets are observed in these sources; the jets can be classified by the radio spectral index and the jet morphology observed: a small-scale, optically thick, steady jet and an extended, optically thin, transient jet~\citep[for a review, see][]{2006csxs.book..381F}. In the hard state, or even the hard-intermediate state, a steady jet is observed~\citep[e.g.,][]{2001MNRAS.322...31F,2001MNRAS.327.1273S,2019ApJ...883..198R}, while during the transition from the the hard-intermediate to the soft-intermediate state, the emission of the steady jet is quenched~\citep{2004MNRAS.355.1105F,2011ApJ...739L..19R}. Around the transition to the soft state, no steady jet is observed but a transient jet is launched, which is bright and consists of discrete relativistic ejecta from the black hole~\citep{1994Natur.371...46M,1995Natur.374..141T,2004ApJ...617.1272C,2012MNRAS.421..468M,2019ApJ...883..198R}.

X-ray emission from black hole binaries (BHBs) shows variability over a broad range of frequencies, and at well defined frequencies~\citep[for a recent review, see][]{2019NewAR..8501524I}. Quasi-periodic oscillations~\citep[QPOs;][]{1989ASIC..262...27V,2000MNRAS.318..361N,2002ApJ...572..392B} are the narrow peaks in the power density spectrum (PDS). The low-frequency QPOs (LFQPOs) in BHBs can be classified into three classes, namely type A, B, and C, based on the shape of the broadband noise in the PDS, the root mean square (rms) amplitude and phase lags of the QPOs, and the spectral state of the source~\citep{2005ApJ...629..403C,2018ApJ...866..122H}. High-frequency QPOs, with frequency up to $\sim$ 350~Hz, in black holes X-ray binaries are less common~\citep[e.g.,][]{1997ApJ...482..993M,1999ApJ...522..397R,2013MNRAS.435.2132M}. The dynamic origin of LFQPOs is explained mainly by geometric effects or instabilities in the accretion flow~\citep{1998ApJ...492L..59S,1999A&A...349.1003T,2009MNRAS.397L.101I,2018ApJ...858...82Y,2021NatAs...5...94M}. The radiative properties of QPOs, rms amplitude and phase lags, provide extra information. The 0.1--10 Hz integrated rms amplitude, for instance, is an indicator of accretion regimes in black hole transients~\citep{2011MNRAS.410..679M}. Type-C QPOs often appear when the source is in the low-hard state and the hard-intermediate state, with central frequency in the range $\sim 0.1$--$15$~Hz and associated strong broadband noise in the PDS~\citep{2005ApJ...629..403C}. The rms amplitude of the type-C QPOs can increase with energy, and reach $\sim$ 15\% at 100 keV~\citep[e.g.,][]{2018ApJ...866..122H,2021ApJ...919...92B}, indicating that the radiative mechanism of the type-C QPO has to be related to the corona.

The phase lags are the phase difference of two light curves at different energies in the same frequency range~\citep{1999ApJ...510..874N}. Hard (positive) lags are believed to be due to disc fluctuations that propagate through the corona~\citep{1988Natur.336..450M}, while soft (negative) lags may come from the reprocessing of the hard photons in the disc when the corona illuminates the disc~\citep{2001ApJ...549L.229L}. The rms spectrum and the phase-lag spectrum of the LFQPOs can indicate how the disc and the Comptonization region influence the evolution of the radiative properties of the LFQPOs~\citep[e.g.,][]{2000ApJ...541..883R,2005A&A...440..207B,2020MNRAS.494.1375Z,2020JHEAp..25...29K}.

Although the corona around a black hole is widely believed to consist of hot electrons with temperatures up to $\sim$ 100~keV that can give rise to the Comptonized spectrum~\citep{1996MNRAS.283..193Z,1999MNRAS.309..561Z}, the geometry of the corona is still under debate~\citep{1979ApJ...229..318G,1991ApJ...380L..51H,2018A&A...614A..79P}. It is therefore of great significance to determine the geometry of the Comptonization region. Models proposed to constrain the properties of the corona describe either time-averaged energy spectrum or the X-ray variability. For instance, \citet{2014ApJ...782...76G} proposed a time-averaged reflection model called \texttt{relxill} that can measure the height of the corona assuming a lamppost geometry. Using this model~\citet{2021NatCo..12.1025Y} provided information about how the corona moves. \citet{2019MNRAS.488..324I} presented a relativistic transfer function model, \texttt{reltrans}, that calculates the X-ray time delays and energy shifts based on reverberation mapping. Both these models have been applied to observations~\citep[e.g.,][]{2018ApJ...852L..34X,2021ApJ...910L...3W}, however, neither the time-averaged reflection model nor the reverberation mapping can explain the radiative properties of the QPOs.

Recently, \citet{2020MNRAS.492.1399K} developed a time-dependent Comptonization model based on the idea proposed by~\citet{1998MNRAS.299..479L} and~\citet{2014MNRAS.445.2818K}. The model of~\citet{2020MNRAS.492.1399K} assumes the QPO to be a sinusoidal coherent oscillation of the Comptonized X-ray spectrum. The oscillation comes from the temperature of the corona, the inner disc, and the external heating source which is necessary to keep the thermal equilibrium of the system. In this model, the corona is assumed to be spherically symmetric and partially covering the accretion disc, while blackbody seed photons from the accretion disc are Compton up-scattered in the corona, leading to hard lags. Since the corona covers the disc to some extent, a fraction of the hard photons in the corona can impinge back onto the accretion disc, resulting in reprocessing of the hard photons and soft lags. The fraction of the feedback photons to the Comptonized photons, ranging from 0 to 1, is called the feedback fraction. The steady-state version of this model describes the Comptonized continuum in the time-averaged energy spectrum, similar, for instance, to \texttt{nthcomp}~\citep{1996MNRAS.283..193Z,1999MNRAS.309..561Z}, while the time-dependent model can fit simultaneously the fractional rms and phase-lag spectra. The latest version (Bellavita et al., 2022) of the model~\citep{2020MNRAS.492.1399K} incorporates a disc blackbody as the soft-photon source so the steady-state version of this model is the same as \texttt{nthcomp}. \citet{2020MNRAS.492.1399K} first applied the model to the kilohertz QPOs in the neutron star 4U~1636$-$53. Later on this model was further applied to the LFQPOs of BHBs: \citet{2021MNRAS.503.5522K} measured a variable corona in \grs as the frequency of the type-C QPO changes using \rxte data; \citet{2021MNRAS.501.3173G} applied a dual Comptonization model to the type-B QPO of MAXI~J1348+603 with \textit{NICER}, showing that there may be two Comptonization regions near the black hole where the outer region can be the jet. Here we use this time-dependent Comptonization model to study the evolution of the Comptonization region of \maxi through type-C QPOs with \textit{Insight}-HXMT.

\maxi is a new X-ray transient discovered independently by \textit{MAXI}/GSC~\citep{2017ATel10708....1N} and \textit{Swift}/BAT~\citep{2017ATel10700....1K} on September 2 2017. Follow-up X-ray and radio observations suggested that this new source is a black hole X-ray binary system~\citep{2017ATel10708....1N,2017ATel10711....1R}. \citet{2018ApJ...860L..28M} analyzed the \textit{NICER} data and proposed that the black hole spin is $0.994\pm 0.002$ and inclination angle is $67.4^{\circ}\pm 0.8^{\circ}$. A broadband X-ray spectral study also indicated that \maxi has a high spin ($>0.84$) and high inclination angle ($\sim 60^{\circ}$)~\citep{2018ApJ...852L..34X}. Type-C QPOs were discovered in the low-hard state of MAXI~J1535$-$571, based on \textit{Swift}, \textit{XMM-Newton}, and \textit{NICER} observations~\citep{2018ApJ...868...71S}. \citet{2019MNRAS.488..720B} showed a tight correlation between the type-C QPOs and the power-law spectral index, $\Gamma$, using \textit{AstroSat} data. \citet{2018ApJ...866..122H} presented a systematic timing study of \maxi with \textit{Insight}-HXMT, classifying different types of LFQPOs and confirming the high inclination angle nature of this system. Using part of the same \hxmt dataset where type-C QPOs were detected, \citet{2020JHEAp..25...29K} analyzed the energy and fractional rms spectra, and gave a picture of a shrinking and hardening corona. The dataset of \maxi with \hxmt provides broadband energy range as well as strong QPO signals that we can utilize to fit with the time-dependent Comptonization model of~\citet{2020MNRAS.492.1399K}.

In this paper, we further explore the \hxmt observations of \maxi during the outburst in 2017/2018. We fit simultaneously the fractional rms, phase-lag, and time-averaged energy spectra of each observation using the time-dependent Comptonization model of~\citet{2020MNRAS.492.1399K}. This paper is organized as follows: In Section~\ref{sec:data} we describe the reduction of the \hxmt data of MAXI~J1535$-$571, and explain how we generate the power density spectra and the cross spectra for different energy bands. We also explain how we simultaneously fit the fractional rms, lags, and time-averaged energy spectra. In Section~\ref{sec:results} we show the temporal evolution of MAXI~J1535$-$571, the dependence of the spectral parameters with QPO frequency and, most importantly, the evolution of the corona. In Section~\ref{sec:discussion} we discuss our results and propose a physical picture to explain the evolution of corona and the possible connection between the corona and the jet.

\section{Observations and Data Analysis}\label{sec:data}

All observations presented here were carried out with \textit{Insight}-HXMT. The payload of \hxmt~\citep{2014SPIE.9144E..21Z} consists of three instruments: the Low Energy X-ray Telescope (LE), the Medium Energy X-ray Telescope (ME), and the High Energy X-ray Telescope (HE). The LE, ME, and HE cover, respectively, the 1--15~keV energy range with 1~ms time resolution, the 5--30~keV energy range with 240~$\mu$s time resolution, and the 20--250~keV energy range with 4~$\mu$s time resolution.

\hxmt triggered Target of Opportunity (ToO) observations of \maxi in the period 2017 September 6--23. We use the \hxmt Data Analysis Software (HXMTDAS) v2.04 to extract and reduce the data. We use the following criteria to establish the good time intervals (GTIs) to filter the data: we require that the offset of the pointing to the source is less than 0.04$^\circ$ and the source is at least 10$^\circ$ above Earth horizon. To filter out high-background intervals, we make a cut on the geometric cutoff rigidity with COR~$>8$~GeV, and we select events that are detected at least 300~s before and after the passage of the satellite through the South Atlantic Anomaly (SAA). We finally select all non-blind detectors to generate the products for LE, ME, and HE, respectively.

We use GHATS~\footnote{\url{http://www.brera.inaf.it/utenti/belloni/GHATS_Package/Home.html}} to compute the Fourier power density spectrum (PDS) of photons in the LE in 1--10~keV energy band, ME in 10--35~keV energy band, and HE in 26--100~keV energy band. We set the time length of each Fast Fourier Transformation (FFT) segment to 25~s and the time resolution to 3~ms so that the lowest frequency of the PDS is 0.04~Hz and the Nyquist frequency is 167~Hz. We average the PDS within one observation, subtract the Poisson level by calculating the average power in the frequency range of $>$ 100~Hz, and normalize the PDS to units of rms$^{2}$ per Hz~\citep{1990A&A...230..103B}. The background count rate used to convert to rms units is estimated with the tools LE(ME/HE)BKGMAP in HXMTDAS. Finally, we apply a logarithmic rebin in frequency to the PDS data such that the size of a bin increases by $\exp(1/100)$ compared to that of the previous one.

We use XSPEC v12.11.1~\citep{1996ASPC..101...17A} to fit the PDS with a model consisting of several Lorentzian functions in the frequency range of 0.5--30.0~Hz. Three Lorentzians represent the low-frequency broadband noise, high-frequency broadband noise, and the QPO. Two extra Lorentzians may be needed to represent the harmonic and the subharmonic of the QPO. Following~\citet{2018ApJ...866..122H} and~\citet{2020JHEAp..25...29K}, the eight observations with type-C QPOs in which all the three instruments are active, are selected for further study. Our observations 1--8 correspond to observations in Table 2 of~\citet{2018ApJ...866..122H} and Table 3 of~\citet{2020JHEAp..25...29K}; all these observations are in the HIMS. Dynamical power spectra of these observations show that the QPO frequency does not change significantly during an observation. We regard the best-fitting model for each observation as a baseline to perform further fitting on the separate energy bands.

To study the rms spectrum of the QPO, we select the following energy bands, LE: 1.0--2.5~keV, 2.5--3.5~keV, 3.5--5.0~keV, and 5.0--10.0~keV, ME: 10.0--12.0~keV, 12.0--14.0~keV, 14.0--17.0~keV, and 17.0--35.0~keV, HE: 26.0--30.3~keV, 30.3--35.5~keV, 35.5--55.0~keV, and 55.0--100.0~keV. We use the background in the corresponding energy band and use the square root of the normalization of the Lorentzian representing the type-C QPOs to calculate the fractional rms amplitude, hereafter rms~\citep{1990A&A...230..103B}. To calculate the error of the rms we include the uncertainties of the background count rates. 

For each observation, we compute Fourier cross spectra for different energy bands to calculate the phase-lag spectrum of the QPOs~\citep{1997ApJ...474L..43V,1999ApJ...510..874N} using the overlapping Good Time Intervals (GTIs) for all the three instruments. The energy bands are the same as for the rms spectra. As we do for the power density spectrum, we use the same time resolution of 3~ms and same time segment of 25~s. We regard the lowest energy band (1.0--2.5~keV) as the reference band. The phase lags of the QPOs are calculated from the average of the real and imaginary parts of the cross spectrum in the frequency range of $\nu_{0}\pm\text{FWHM}/2$, where $\nu_{0}$ is the centroid frequency of the QPOs. A positive lag means that the hard photons lag the soft ones.

\subsection{Fitting the time-averaged, rms, and phase-lag spectra separately}\label{subsec:sep}

We use XSPEC v12.11.1~\citep{1996ASPC..101...17A} to fit the time-averaged, rms, and phase-lag spectra. We adopt the energy bands 2--10 keV (LE), 10--35 keV (ME), and 26-100 keV (HE) for the fitting of the time-averaged spectrum, adding a systematic error of 1\%. Since we assume that the variability at the QPO frequency comes from the corona, we fit the time-averaged and the time-dependent spectra using the same model in which we consider the Comptonization of the seed photons from the accretion disc as the component that drives the variability. We start by fitting the time-averaged spectrum in each observation with the model \texttt{const*tbabs*(diskbb+nthcomp)}, where the multiplicative constant is used to account for cross-calibration uncertainties between three instruments. In the absorption model \texttt{tbabs}, the parameter $N_{\text{H}}$ represents the hydrogen column density, with the cross-section and solar-abundance tables of \citet{1996ApJ...465..487V} and \citet{2000ApJ...542..914W}, respectively. To fit the emission of the accretion disc, we use \texttt{diskbb}~\citep{1984PASJ...36..741M}, which has two parameters: the inner disc temperature, $kT_{\text{in}}$, and a normalization. To fit the hard Comptonized continuum, we use \texttt{nthcomp}~\citep{1996MNRAS.283..193Z,1999MNRAS.309..561Z}, which has the following parameters: the corona temperature, $kT_{\text{e}}$, the X-ray photon index, $\Gamma$, the seed photon temperature, $kT_{\text{s}}$, and a normalization. We assume that the seed photon source is the disc so when performing the fitting the $kT_{\text{in}}$ in \texttt{diskbb} and $kT_{\text{s}}$ in \texttt{nthcomp} are linked.

Fits with the model \texttt{const*tbabs*(diskbb+nthcomp)} show residuals at $\sim$ 6--7~keV and $\sim$ 20--30~keV, suggesting that there may exist a reflection component~\citep[e.g.,][]{2018ApJ...852L..34X,2020MNRAS.492.1947J}. We therefore add a relativistic reflection component, \texttt{relxillCp}~\citep{2014ApJ...782...76G}, to the model that now is \texttt{const*tbabs*(diskbb+nthcomp+relxillCp)}. We link $kT_{\text{e}}$ and $\Gamma$ in \texttt{relxillCp} to $kT_{\text{e}}$ and $\Gamma$ in \texttt{nthcomp} and, as before, $kT_{\text{s}}$ in \texttt{nthcomp} to $kT_{\text{in}}$ in \texttt{diskbb}. In \texttt{relxillCp}, the black hole spin, $a_{*}$, inclination angle, $i$, disc inner radius, $R_{\text{in}}$, logarithm of the ionization parameter, $\log\xi$, iron abundance, $A_{\text{Fe}}$, and normalization of the model are free during the fits, whereas the reflection fraction is fixed at $-1$ since we only consider the reflected emission (the direct emission is given by \texttt{nthcomp}); we also link the emissivity indices $q_{1}$ and $q_{2}$ during the fits.

To fit the rms and phase-lag spectra, we utilize the model of~\citet{2020MNRAS.492.1399K} with a disc blackbody spectrum for the seed-photon source (Bellavita et al. 2022), which we call \texttt{vkompthdk} in this paper. Since both \texttt{vkompthdk} and \texttt{nthcomp} come from the same calculation of the Comptonization spectrum, the time-averaged version of \texttt{vkompthdk} is the same as \texttt{nthcomp}, and both models share the same parameters. However, the time-dependent version of \texttt{vkompthdk} has extra parameters, namely the size of the corona, $L$, feedback fraction, $\eta$, amplitude of the variability of the external heating rate, $\delta\dot{H}_{\text{ext}}$, QPO frequency and a reference lag. The reference lag is an additive parameter that accounts for the fact that the reference band of the lags is arbitrary, and hence the lag spectrum can be shifted up and down arbitrarily~\citep[e.g.,][]{2021MNRAS.501.3173G}.

\subsection{A joint fitting}\label{subsec:joint}

Since our aim is to fit the rms, phase-lag, and time-averaged spectra simultaneously, we use \texttt{vkompthdk} instead of \texttt{nthcomp} to fit the three spectra given that the time-averaged version of \texttt{vkompthdk} is the same as \texttt{nthcomp}. We therefore fit jointly the three spectra in each observation where the energy ranges we use per instrument are 1--10~keV, 10--35~keV, and 26--100~keV for the rms and phase-lag spectra, and 2--10~keV, 10--35~keV, and 26--100~keV for the time-averaged spectra. The model that we use is written as \texttt{const*tbabs*(diskbb+vkompthdk+relxillCp)}. We let $N_{\text{H}}$ and the normalizations of \texttt{diskbb} and \texttt{relxillCp} free to fit the time-averaged spectrum, while we fix all of them at 0 to fit the rms and phase-lag spectra. For \texttt{vkompthdk} the normalization is free when fitting the time-averaged spectrum and fixed at 1 for the rms and phase-lag spectra~\citep[see][]{2021MNRAS.501.3173G}. Besides these parameters, the other parameters that are always free are: $kT_{\text{in}}$ in \texttt{diskbb}, $kT_{\text{e}}$ and $\Gamma$ in the time-averaged version of \texttt{vkompthdk}, where $kT_{\text{s}}$ in \texttt{vkompthdk} is linked to $kT_{\text{in}}$ in \texttt{diskbb}, $q_{1}$, $a_{*}$, $i$, $R_{\text{in}}$, $\log\xi$, and $A_{\text{Fe}}$ in \texttt{relxillCp}, where $kT_{\text{e}}$ and $\Gamma$ in \texttt{relxillCp} are linked to the same parameters in \texttt{vkompthdk}. The reflection fraction is fixed to $-1$ so that \texttt{relxillCp} only provides the reflected component, while \texttt{vkompthdk} provides the direct emission from the hard Comptonization component. When we use the time-dependent version of \texttt{vkompthdk} to fit the rms and phase-lag spectra, the QPO frequency is fixed at the frequency of the QPO in the PDS in each observation, while $L$, $\eta$, $\delta\dot{H}_{\text{ext}}$ and the reference lag, which act only on the phase-lag spectrum as mentioned in~\ref{subsec:sep}, are free.

We fit the data from the three instruments, i.e.\ LE, ME, and HE, simultaneously, so in each observation we have three time-averaged, three rms and three phase-lag spectra, respectively, and the parameters in the model of each component are linked across different instruments. Initially the spin, $a_{*}$, and the inclination angle, $i$, in the reflection model are free to change from one observation to the other. However, we notice that these parameters are high and more or less constant when the QPO frequencies are low, but drop when the QPO frequencies are high. \citet{2018ApJ...860L..28M} and~\citet{2018ApJ...852L..34X} suggested a high spin ($>0.84$) and a high inclination angle ($>60^{\circ}$) for MAXI~J1535$-$571. Since in principle these parameters should not change from one observation to the other, we fix the spin and inclination angle at the average values of all our fits, i.e.\ 0.998 and 60.2$^{\circ}$, respectively. In the fits the hydrogen column density $N_{\text{H}}$ is also fixed at the average value obtained from the fits of the eight observations.

\section{Results}\label{sec:results}

\subsection{The light curve and HID of \maxi}

\begin{figure*}
    \includegraphics[width=0.49\textwidth]{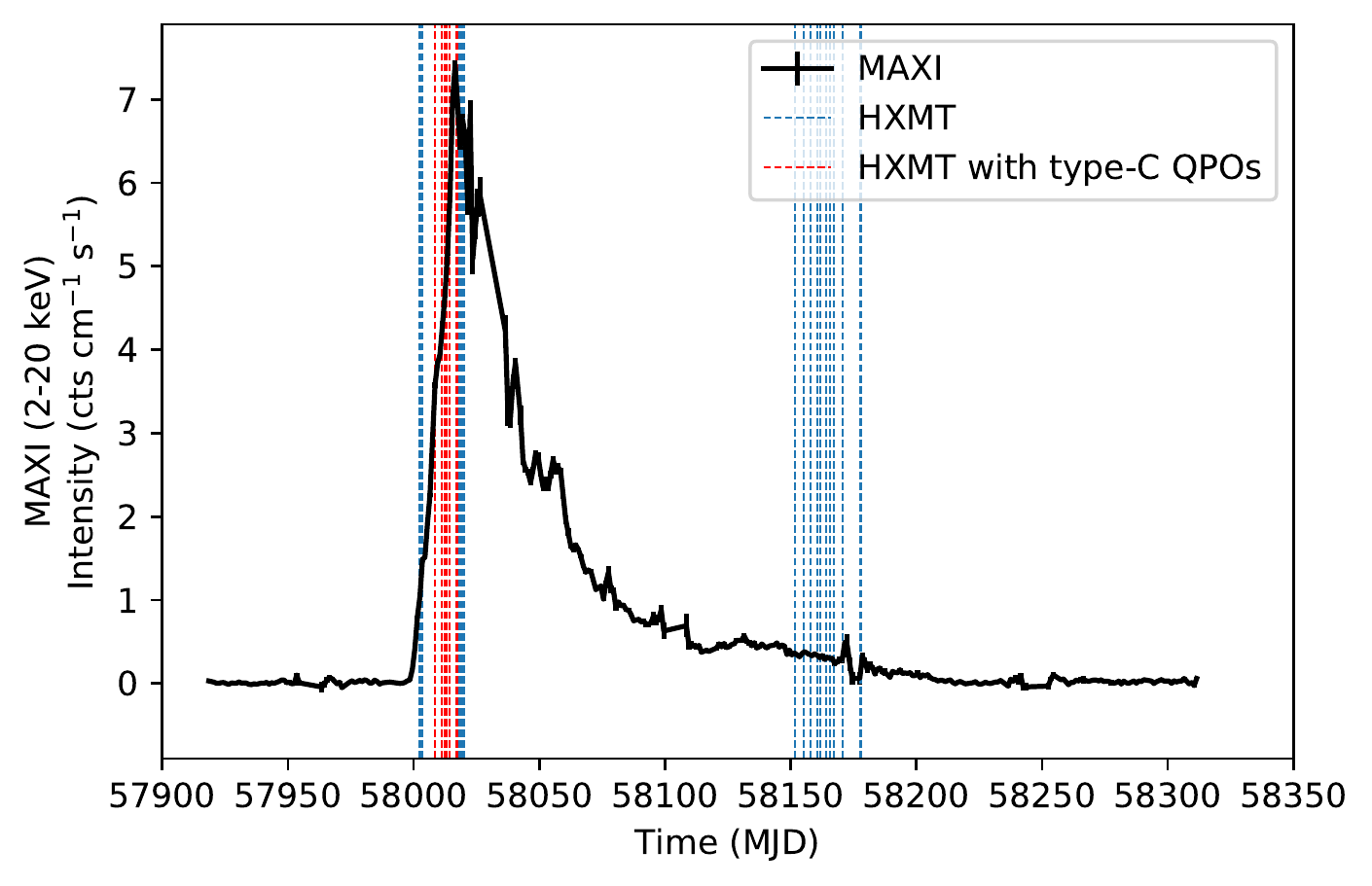}
    \includegraphics[width=0.49\textwidth]{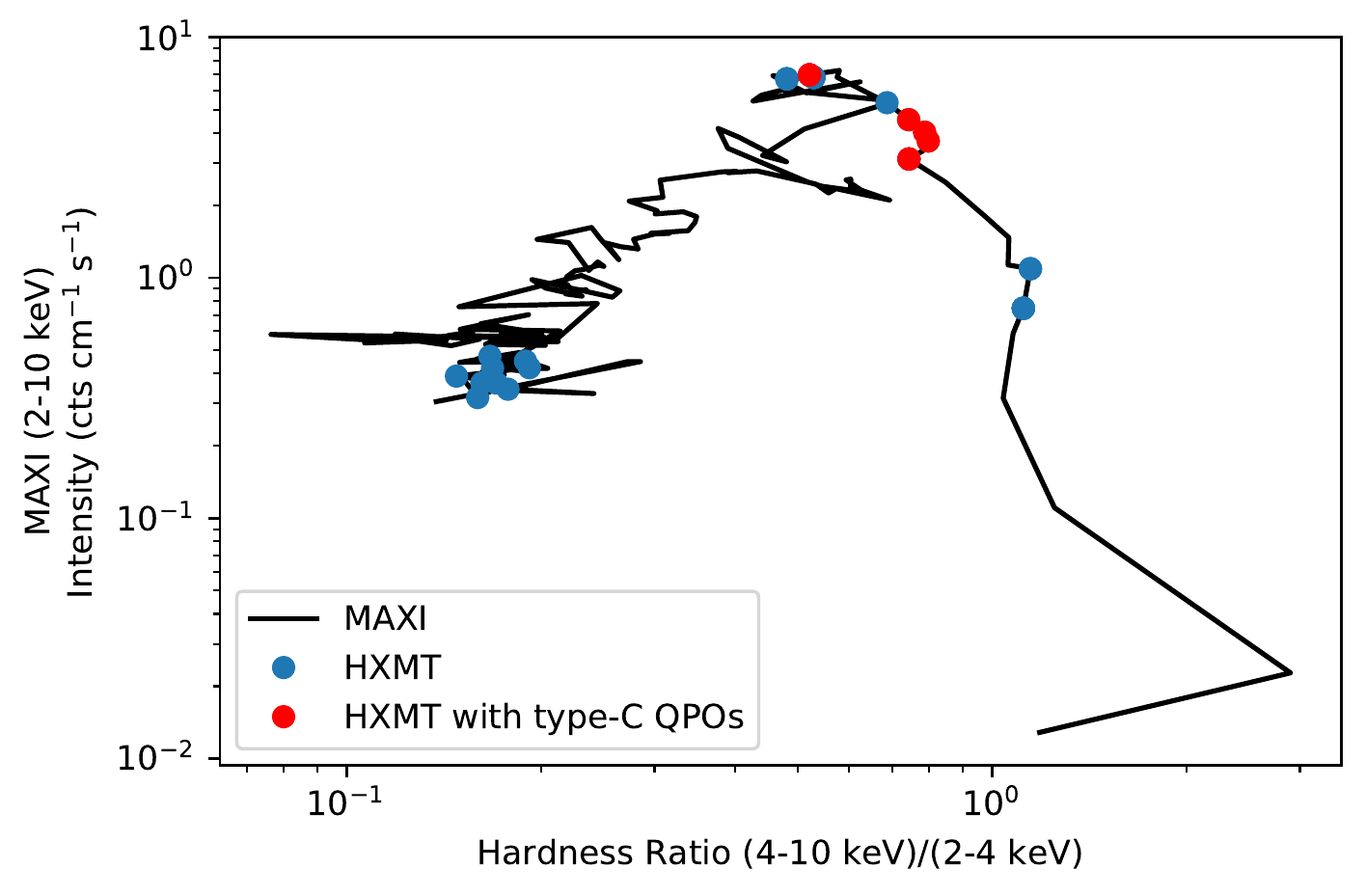}
    \caption{Left panel: MAXI light curve of \maxi in black with the time of all the \hxmt observations indicated by the vertical dashed lines. The selected MJD range of MAXI observations is from 57918 to 58312. The red dashed lines represent the observations that we analyze with type-C QPOs in the intermediate state. Right panel: Hardness intensity diagram (HID) of \maxi using MAXI data in black. The selected MJD range of MAXI observations is from 57997 to 58174. The blue and red points indicate the simultaneous \hxmt observations. The red points represent the observations that we analyze with type-C QPOs in the intermediate state. Note that some of the red points overlap each other. There are 3 red points in the left group and 5 red points in the right group.}
    \label{fig:lc}
\end{figure*}

In the left panel of Figure~\ref{fig:lc} we show the MAXI light curve of \maxi in the 2--20~keV band during the 2017/2018 outburst, from MJD 57918 to MJD 58312. The X-ray emission of \maxi is undetectable until $\sim$ MJD 58000. From that date onwards the source displays a rise up to a maximum intensity of $\sim$ 7 counts cm$^{-2}$ s$^{-1}$ on MJD 58016, and after that it decays gradually. The vertical dashed lines in Figure~\ref{fig:lc} show that the \hxmt observations of \maxi only cover part of the full outburst, especially the rising phase and the end of the outburst. \citet{2018ApJ...866..122H} reported the type-C QPOs detected by \hxmt in the intermediate state of the source. Those observations are marked with red dashed lines in Figure~\ref{fig:lc}. Radio observations of \maxi start from MJD 58000 until just after MJD 58050, covering the whole intermediate states of the source~\citep{2019ApJ...883..198R,2021PASA...38...45C}.

The MAXI hardness intensity diagram (HID) of \maxi in the right panel of Figure~\ref{fig:lc} shows that the source traces the characteristic `q' path, which is typical for other black hole transients~\citep{2004MNRAS.355.1105F}. During the transition to the high-soft state, the hardness ratio drops as the count rate increases. \hxmt observations after the intermediate state show that the source goes into a low-luminosity state. We also plot in red the eight observations with type-C QPOs that we analyze in detail in this paper. As shown in Table~\ref{tab:qpo}, according to the classification by~\citet{2018ApJ...866..122H}, observations 1--5 were in the hard-intermediate state (HIMS), while observations 6--8 were in the soft-intermediate state (SIMS). We note, however, that the presence of a type-C QPO in those observations, also classified as such by~\citet{2018ApJ...866..122H}, indicates that the source was instead in the HIMS. \citet{2018ApJ...866..122H} found a type-B QPO, which is part of the definition of the SIMS, in observation 701, between the first five and the last three observations in our sample. This indicates that during the first five observations in our sample the source was in the HIMS, it went briefly into the SIMS in observation 701 of~\citet[][this observation is not included in our sample]{2018ApJ...866..122H}, and then went back to the HIMS in observations 901--903 in our sample. Note that in the right panel of Figure~\ref{fig:lc} a blue point (obs ID P011453500601) between the red points corresponds also to an observation in the intermediate state with type-C QPOs. However we exclude this observation since the GTIs of LE, ME, and HE do not overlap and thus we cannot compute the phase-lag spectrum of this observation.

\subsection{Joint fits of the rms, phase-lag, and time-averaged spectra}

\begin{figure*}
    \includegraphics[width=0.49\textwidth,trim= 40 10 90 10]{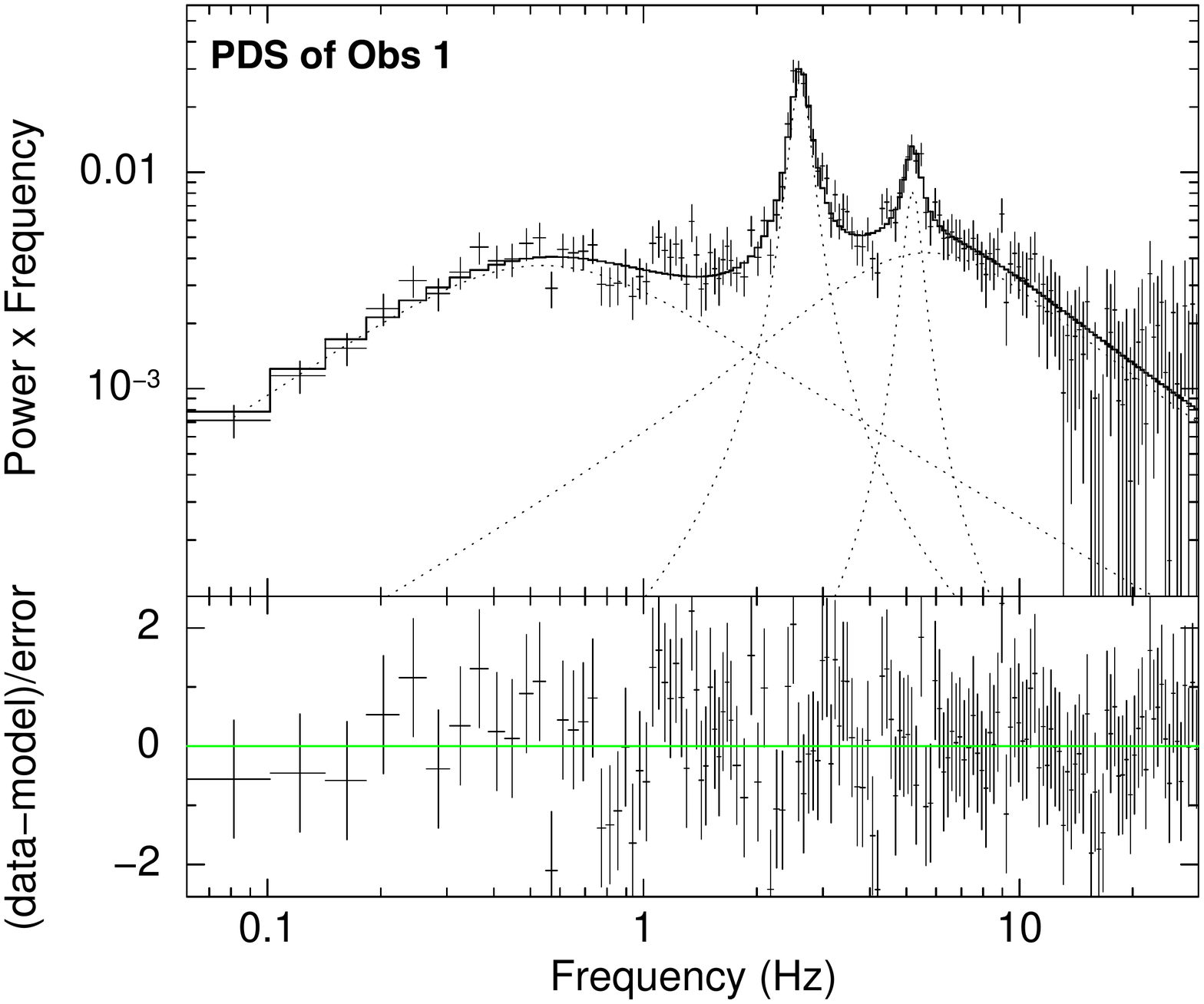}
    \includegraphics[width=0.49\textwidth,trim= 40 10 90 10]{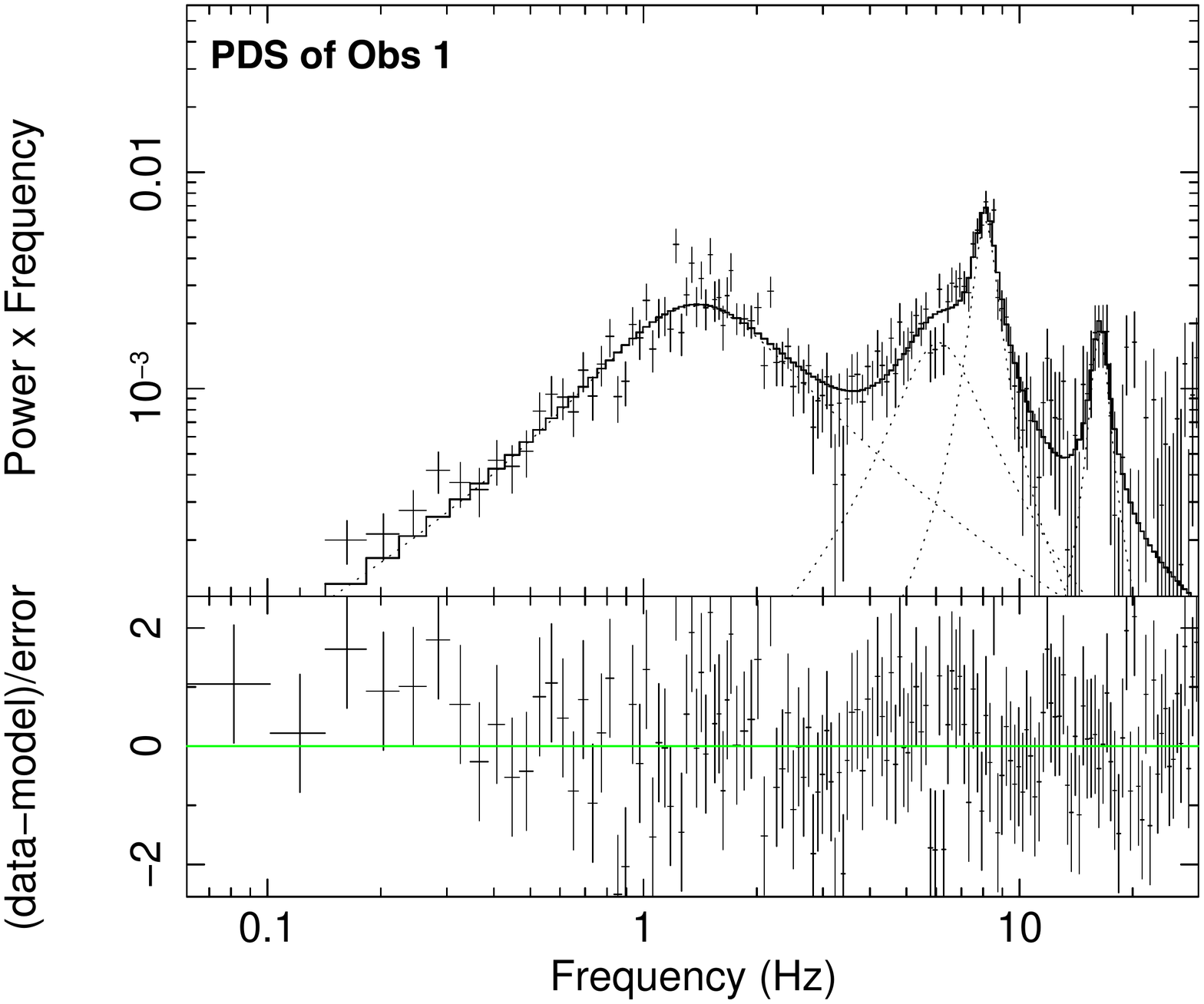}
    \caption{Two representative PDS of \maxi using the 1--10~keV LE data of \hxmt. Left panel: The PDS (top) and the best-fit residuals (bottom) of Obs 1. Right panel: The PDS (top) and the best-fit residuals (bottom) of Obs 8.}
    \label{fig:pds}
\end{figure*}

\begin{figure*}
    \includegraphics[width=0.49\textwidth,trim= 30 25 50 60]{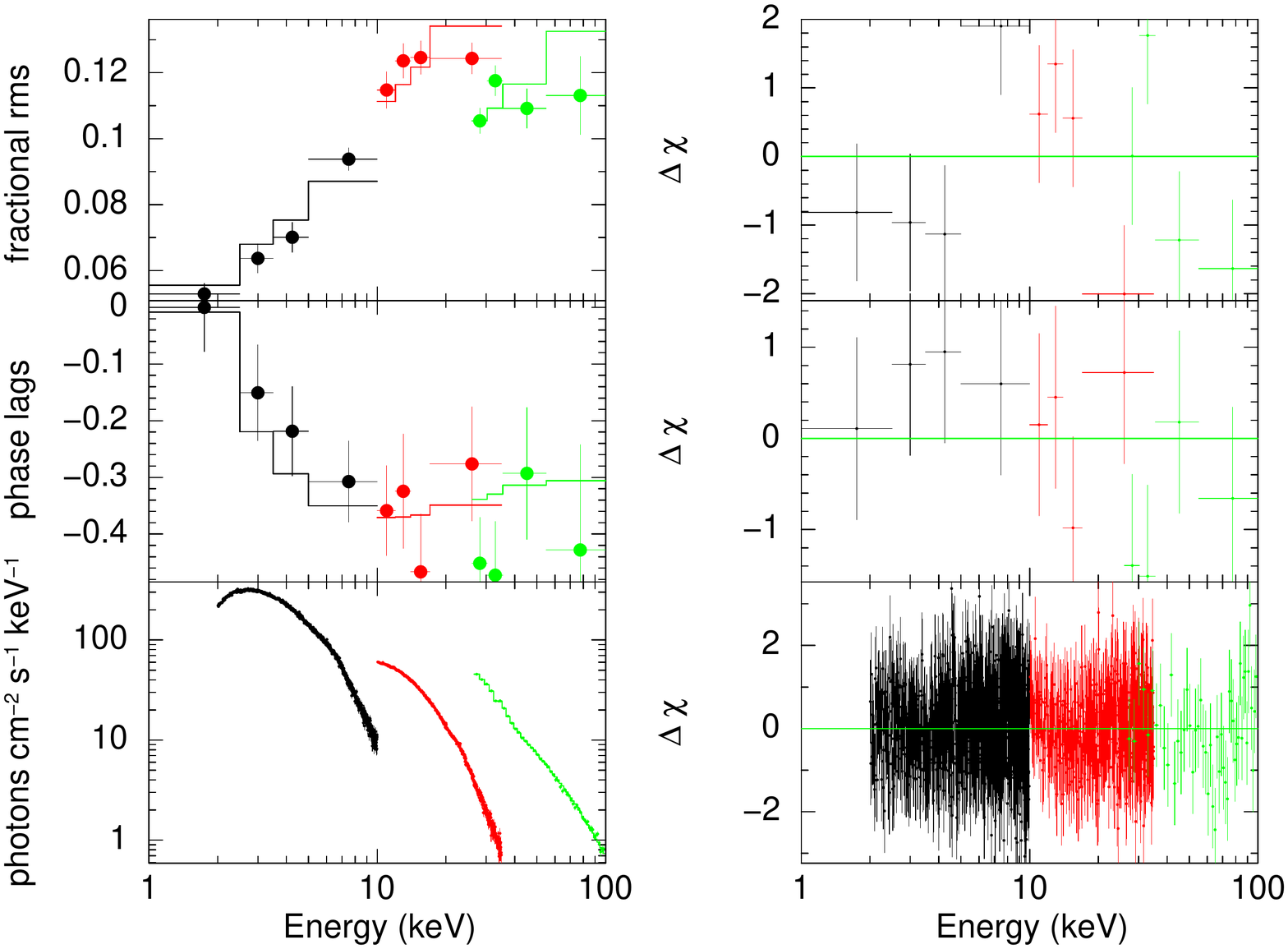}
    \includegraphics[width=0.49\textwidth,trim= 30 25 50 60]{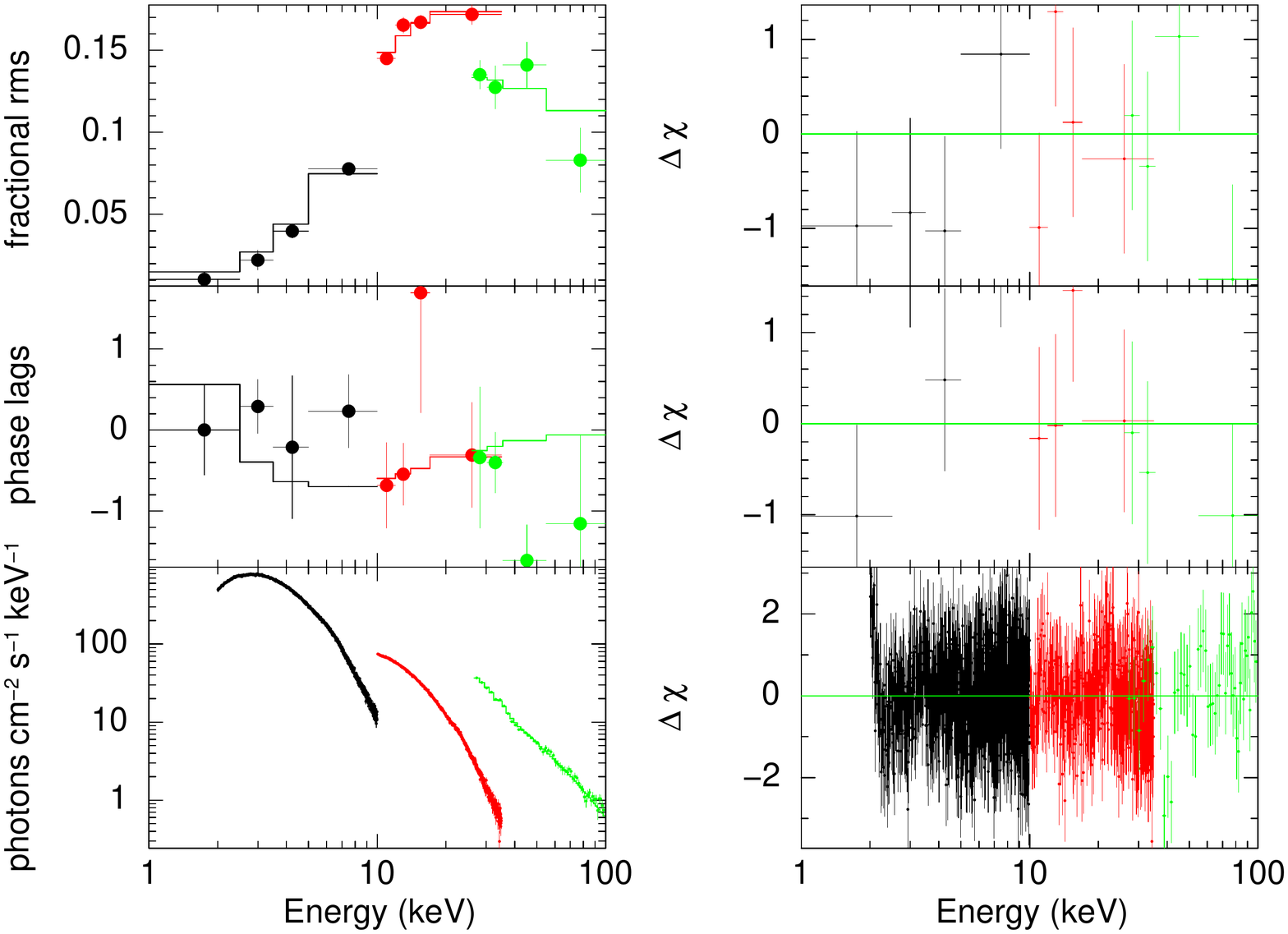}
    \caption{Two representative rms (top panel), phase-lag (middle panel), and time-averaged (bottom panel) spectra with joint fitting of \maxi using \textit{Insight}-HXMT. The first panel shows the spectra for Obs 1 (QPO frequency 2.6~Hz), while the second panel shows the residuals with respect to the best-fitting model. The third and fourth panels show the same for Obs 8 (QPO frequency 8.1~Hz). The black, red, and green colors indicate the LE, ME, and HE data of \textit{Insight}-HXMT, respectively.}
    \label{fig:spectra}
\end{figure*}

\begin{table}
    \caption{Observations and QPOs of MAXI~J1535$-$571. The error bar indicates the 90\% confidence level. Each observation ID has a prefix of P011453500.}
    \centering
    \begin{tabularx}{\columnwidth}{ccccc} 
        \hline
        & Obs IDs & MJD & QPO Frequency (Hz) & FWHM \\
        \hline
        Obs 1 & 144 & 58008.44 & $2.61\pm 0.02$ & $0.32\pm 0.06$ \\
        Obs 2 & 145 & 58008.58 & $2.58\pm 0.02$ & $0.63\pm 0.05$ \\
        Obs 3 & 301 & 58011.20 & $2.12\pm 0.02$ & $0.35\pm 0.10$ \\
        Obs 4 & 401 & 58012.26 & $2.73\pm 0.02$ & $0.29\pm 0.06$ \\
        Obs 5 & 501 & 58013.26 & $3.34\pm 0.02$ & $0.39\pm 0.05$ \\
        Obs 6 & 901 & 58017.10 & $8.88\pm 0.08$ & $0.93\pm 0.19$ \\
        Obs 7 & 902 & 58017.25 & $9.02\pm 0.06$ & $1.04\pm 0.19$ \\
        Obs 8 & 903 & 58017.39 & $8.08\pm 0.11$ & $1.04\pm 0.34$ \\
        \hline
    \end{tabularx}\label{tab:qpo}
\end{table}

\begin{table*}
    \caption{Best-fitting spectral parameters for the model \texttt{tbabs*(diskbb+vkompthdk+relxillCp)} used to fit the data of MAXI~J1535$-$571. The error bars indicate the 90\% confidence level.}\label{tab:par}
    \centering
    \begin{threeparttable}
        \begin{tabularx}{\textwidth}{ccccccccc} 
        \hline\hline
                                   & Obs 1 & Obs 2 & Obs 3 & Obs 4 & Obs 5 & Obs 6 & Obs 7 & Obs 8 \\
        \hline
        $N_{\text{H}}$ (10$^{22}$) & $5.553^{*}$                        & $5.553^{*}$                        & $5.553^{*}$                           & $5.553^{*}$                        & $5.553^{*}$                      & $5.553^{*}$                       & $5.553^{*}$                        & $5.553^{*}$                        \\
        $kT_{\text{s}}$ (keV)      & $0.27\pm 0.03$          & $0.23\pm 0.02$          & $0.24\pm 0.01$             & $0.21\pm 0.05$          & $0.23\pm 0.05$        & $0.78\pm 0.06$         & $0.70\pm 0.05$          & $0.71\pm 0.04$          \\
        Norm$_{\text{disc}}$ ($10^{3}$)       & $1700^{+2500}_{-1000}$     & $5900^{+9800}_{-3100}$    & $5500^{+1700}_{-1300}$       & $16000^{+192000}_{-10000}$ & $3600^{+33100}_{-3100}$ & $7.4^{+2.1}_{-1.6}$            & $13\pm 3$               & $8.3^{+1.5}_{-2.6}$             \\
        $\Gamma$                   & $2.49\pm 0.03$          & $2.49\pm 0.01$          & $2.43\pm 0.02$             & $2.59\pm 0.02$          & $2.60\pm 0.02$        & $2.83^{+0.02}_{-0.05}$         & $2.88\pm 0.02$          & $2.80\pm 0.02$          \\
        $kT_\text{e}$ (keV)        & $32\pm 3$         & $34\pm 2$         & $34\pm 2$            & $44^{+8}_{-4}$         & $39\pm 3$       & $202^{+44}_{-53}$     & $250^{}_{-40}$                & $250^{}_{-48}$    \\
        $q_{1}$                    & $8.1^{+1.8}_{-1.3}$          & $9.7^{+(P)}_{-1.1}$         & $6.0\pm 0.3$             & $6.1\pm 0.2$          & $5.9\pm 0.2$        & $8.4\pm 0.4$         & $8.9\pm 0.2$          & $8.7\pm 0.3$          \\
        $a_{*}$                    & $0.998^{*}$                        & $0.998^{*}$                        & $0.998^{*}$                           & $0.998^{*}$                        & $0.998^{*}$                      & $0.998^{*}$                       & $0.998^{*}$                        & $0.998^{*}$                        \\
        $i$ ($^{\circ}$)           & $60.2^{*}$                         & $60.2^{*}$                         & $60.2^{*}$                            & $60.2^{*}$                         & $60.2^{*}$                       & $60.2^{*}$                        & $60.2^{*}$                         & $60.2^{*}$                         \\
        $R_{\text{in}}$ (ISCO)     & $1.64\pm 0.06$          & $1.66\pm 0.03$          & $1.00^{+0.06}_{}$            & $1.00^{+0.28}_{}$          & $1.00^{+0.06}_{}$        & $1.08\pm 0.02$         & $1.08\pm 0.01$         & $1.07\pm 0.02$          \\
        $\log\xi$                  & $2.83^{+0.26}_{-0.07}$          & $2.89\pm 0.07$          & $2.86^{+0.07}_{-0.11}$             & $2.7\pm 0.1$          & $3.09^{+0.08}_{-0.05}$        & $4.70^{}_{-0.04}$        & $4.70^{}_{-0.03}$          & $4.6^{+(P)}_{-0.1}$         \\
        $A_{\text{Fe}}$            & $0.95^{+0.75}_{-0.12}$          & $0.58\pm 0.06$          & $0.51^{+0.11}_{-0.01}$             & $0.72\pm 0.09$          & $0.88^{+0.08}_{-0.11}$        & $2.1\pm 0.3$         & $3.2^{+1.0}_{-0.3}$          & $2.0\pm 0.2$          \\
        Norm$_{\text{refl}}$       & $0.53\pm 0.09$          & $0.60\pm 0.05$          & $1.1\pm 0.1$             & $1.8\pm 0.3$          & $1.7\pm 0.1$        & $3.3\pm 0.2$         & $3.0\pm 0.2$          & $3.2\pm 0.2$          \\
        $L$ ($10^{3}$ km)                   & $5.1\pm 2.0$ & $4.4\pm 1.2$ & $3.4\pm 0.4$ & $3.0\pm 0.9$   & $3.2^{+0.3}_{-0.6}$ & $5.5^{+1.2}_{-0.9}$ & $4.6\pm 0.7$ & $6.9^{+1.5}_{-1.2}$ \\
        $\eta$                     & $0.85^{+0.13}_{-0.10}$          & $0.74\pm 0.06$          & $0.77\pm 0.05$            & $0.90\pm 0.05$          & $0.96^{+(P)}_{-0.04}$       & $0.54^{+0.18}_{-0.15}$         & $0.48\pm 0.16$          & $0.78^{+(P)}_{-0.26}$         \\
        $\delta\dot{H}_{\text{ext}}$     & $0.085\pm 0.007$          & $0.097\pm 0.004$          & $0.0817\pm 0.003$             & $0.072\pm 0.004$          & $0.077\pm 0.007$        & $0.089\pm 0.015$         & $0.095\pm 0.006$          & $0.13\pm 0.02$          \\
        Norm$_{\text{Comp}}$       & $0.23^{+0.20}_{-0.12}$          & $0.53^{+0.35}_{-0.19}$          & $0.51\pm 0.06$             & $2.5^{+1.9}_{-1.3}$          & $1.5\pm 1.0$        & $0.014^{+0.013}_{-0.006}$         & $0.045\pm 0.021$          & $0.025^{+0.015}_{-0.009}$          \\
        \hline
        $\chi^{2}/\nu$ & $1194.8/1426$ & $1284.8/1426$ & $1415.4/1426$ & $1358.0/1426$ & $1484.7/1426$ & $1404.8/1426$ & $1381.6/1426$ & $1330.6/1426$ \\
        \hline
        \end{tabularx}
        \begin{tablenotes}
            \item[1] The letter $(P)$ indicates that the positive 90\% error of the parameter pegged at the upper boundary.
        \end{tablenotes}
    \end{threeparttable}
\end{table*}

In Table~\ref{tab:qpo} we show the \hxmt observation IDs of \maxi with type-C QPOs; the QPO frequency is in the range of 2.12--9.02~Hz, and the source is in the intermediate state. Our classification of the QPOs as type-C QPOs is consistent with the classification in these observations by~\citet{2018ApJ...866..122H} although, as we mentioned above, \citet{2018ApJ...866..122H} misclassified the state of the source in the last three observations presented here. From MJD 58008 to MJD 58013, when the type-C QPO frequency decreases from 2.61 Hz to 2.12 Hz and then increases to 3.34 Hz, the source is in the HIMS. According to~\citet{2018ApJ...866..122H}, on MJD 58015 a type-B QPO is detected, which indicates that on this date the source is in the SIMS. (Notice that we do not include this observation and this QPO in our analysis.) Finally, the three observations on MJD 58017 contain again a type-C QPO with a frequency that shows a fast increase from 3.34 Hz to around 9 Hz. The presence of the type-C QPO indicates that, after a short excursion, the source is again in the HIMS. The variable frequency of the type-C QPOs in the HIMS is consistent with the measurements of~\citet{2019MNRAS.488..720B} who had a much better time sampling. The left panel of Figure~\ref{fig:pds} shows the PDS of observation 1, while the right panel shows the PDS of observation 8. Both panels display a strong signal of the type-C QPOs and the harmonics. In the first and third panels of Figure~\ref{fig:spectra} we plot the rms, phase-lag, and time-averaged spectra when the QPO frequencies are 2.61~Hz and 8.08~Hz, while the second and fourth panels show the residuals to the best-fitting model. When the QPO frequency is 2.61~Hz, the rms first increases with energy and gradually flattens above 10~keV, while the phase lags are always negative (soft) and their magnitude increases with energy. When the QPO frequency is 8.08~Hz, the rms increases with energy, gradually flattens above 10~keV, and drops slightly above 30~keV, while the phase lags are negative and more or less constant with energy, but with larger error bars than in the case when the QPO frequency is at 2.61~Hz. The fact that the lags are soft indicates that soft photons emitted by the disc might be largely due to hard photons from the illuminating corona after being reprocessed~\citep{2001ApJ...549L.229L,2020MNRAS.492.1399K}.

We fit the rms, phase-lag, time-averaged spectra jointly using the method described in Section~\ref{subsec:joint}. As an example, Figure~\ref{fig:spectra} shows the best-fitting model to the rms, phase-lag, and time-averaged spectra when the QPO frequency is 2.61~Hz (first panel, with residuals in the second panel) and 8.08~Hz (third panel, with residuals in the fourth panel).

In observations 1--5, when the QPO frequency is in the range of 2.12--3.34~Hz, the power-law index of the Comptonized component in the time-averaged spectrum is in the range of 2.4--2.6 and the disc temperature is around 0.3~keV, while in the observations 6--8, when the QPO frequency is about 9~Hz the power law is steeper, with $\Gamma\sim$ 2.8, and the temperature of the disc increases to $\sim$ 1.0~keV (see Figure~\ref{fig:evolution} and Table~\ref{tab:par}). In observations 1--5 the temperature of the corona is $kT_{\text{e}}\sim$ 40~keV, whereas in observations 6--8 $kT_{\text{e}}$ changes quickly and reaches the upper limit of 250~keV that we set in the fits. The change of the values of spectral parameters in the time-averaged spectrum indicates that \maxi may be about to undergo a state transition from the HIMS to the SIMS. The normalization of the disc component, Norm$_{\text{disc}}$, decreases as the temperature of the disc, $kT_{\text{s}}$, increases; this anti-correlation is consistent with the findings of~\citet{2018MNRAS.480.4443T}. The emissivity index, $q_{1}$, in the \texttt{relxillCp} component is usually larger than 6, indicating that the corona is close to the black hole~\citep{2014ApJ...782...76G,2018ApJ...864...25G} or the radial dependence of the disc ionization state~\citep{2012A&A...545A.106S}. The best-fitting values of the inner radius of the accretion disc, $R_{\text{in}}$, are consistent with the innermost stable circular orbit (ISCO), which suggests that the inner disc is stable from MJD 58008 to MJD 58017. In the HIMS, the logarithm of the disc ionization parameter is around 3 and the iron abundance is less than that of the Sun; in the late stage of the HIMS, the disc is more ionized with the ionization parameter increasing by an order of the magnitude and the best-fitting iron abundance is around 2 times the solar abundance.

\subsection{The evolution of the geometry of the corona}

\begin{figure*}
    \includegraphics[width=0.49\textwidth]{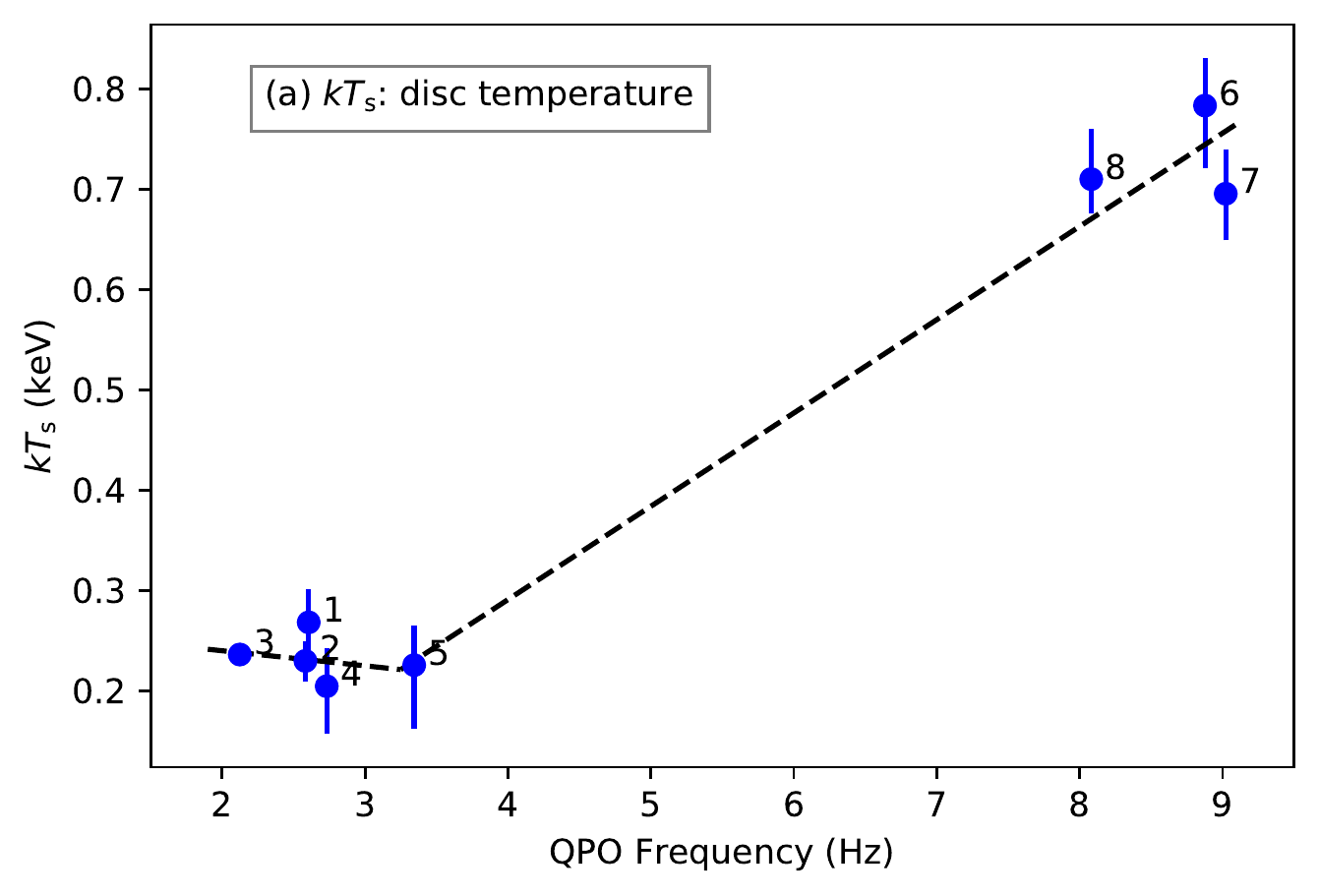}
    \includegraphics[width=0.49\textwidth]{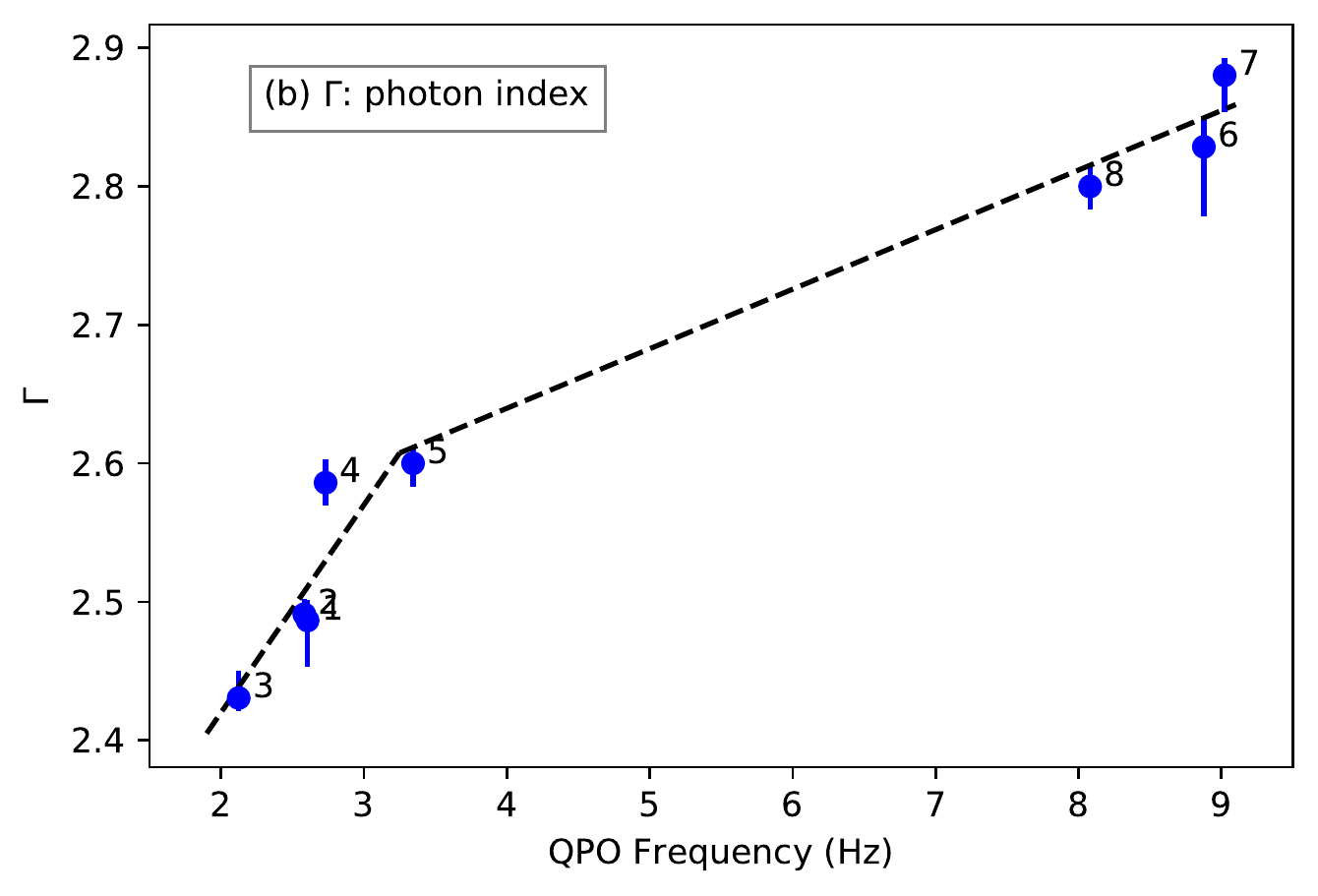}
    \includegraphics[width=0.49\textwidth]{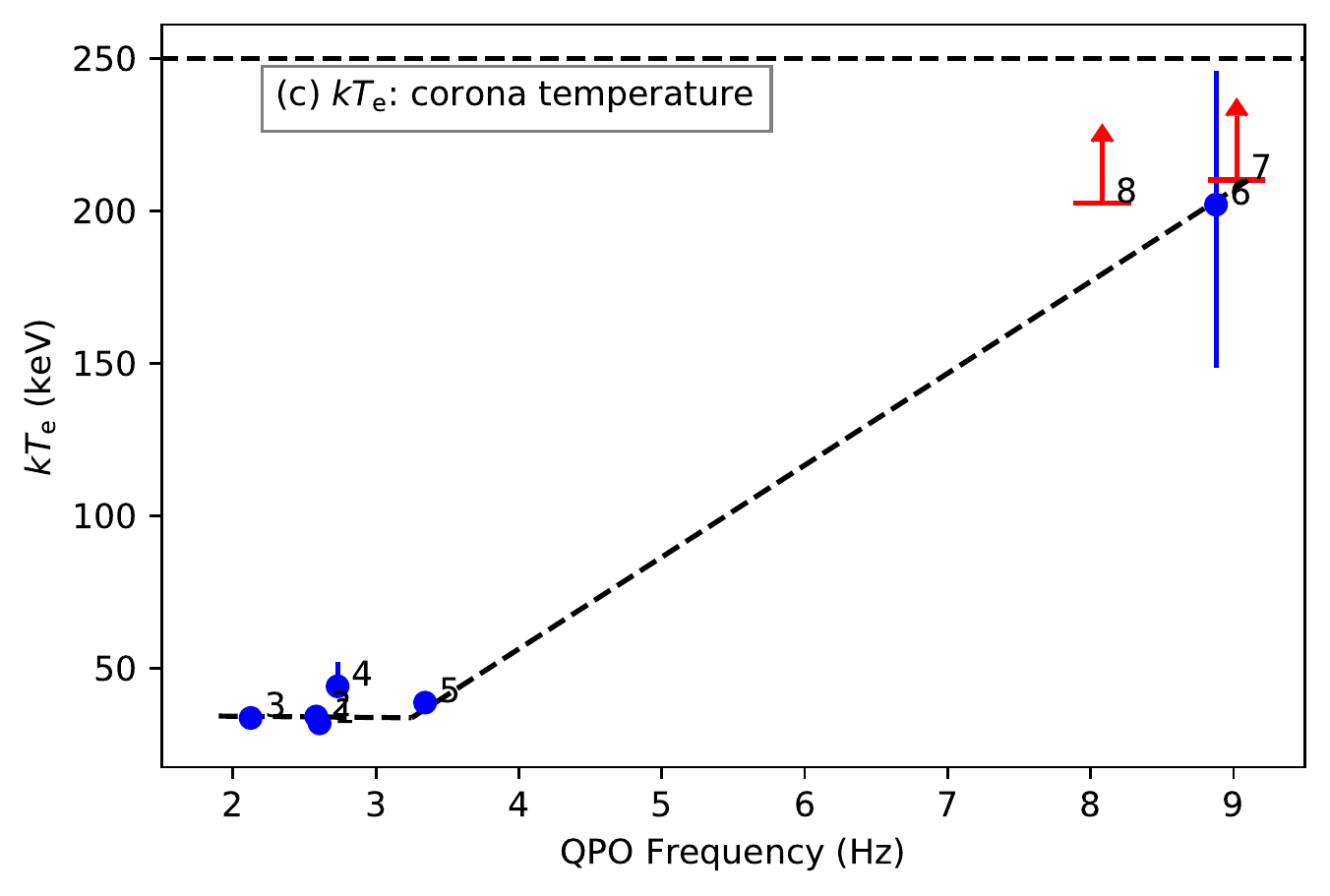}
    \includegraphics[width=0.49\textwidth]{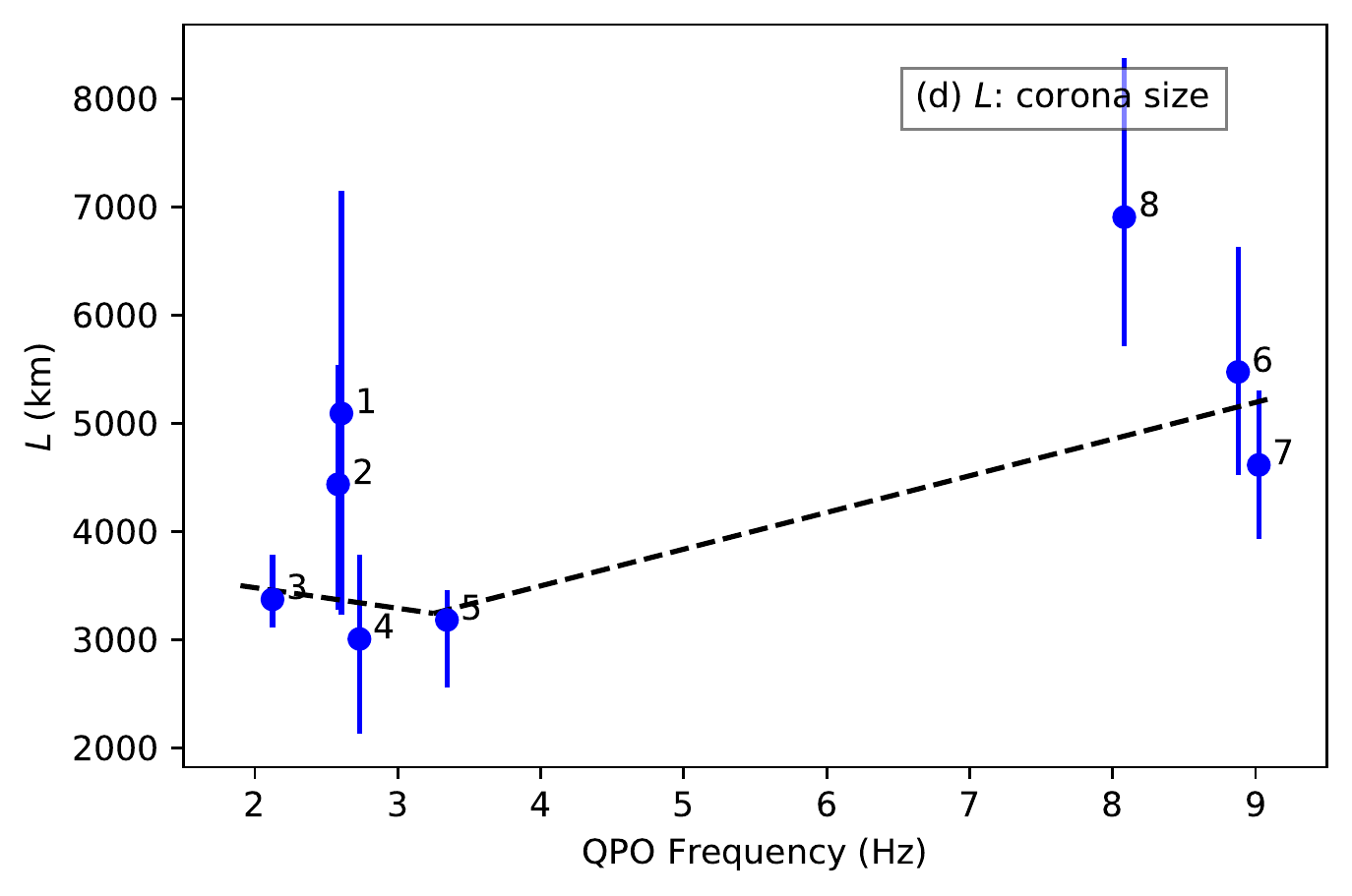}
    \includegraphics[width=0.49\textwidth]{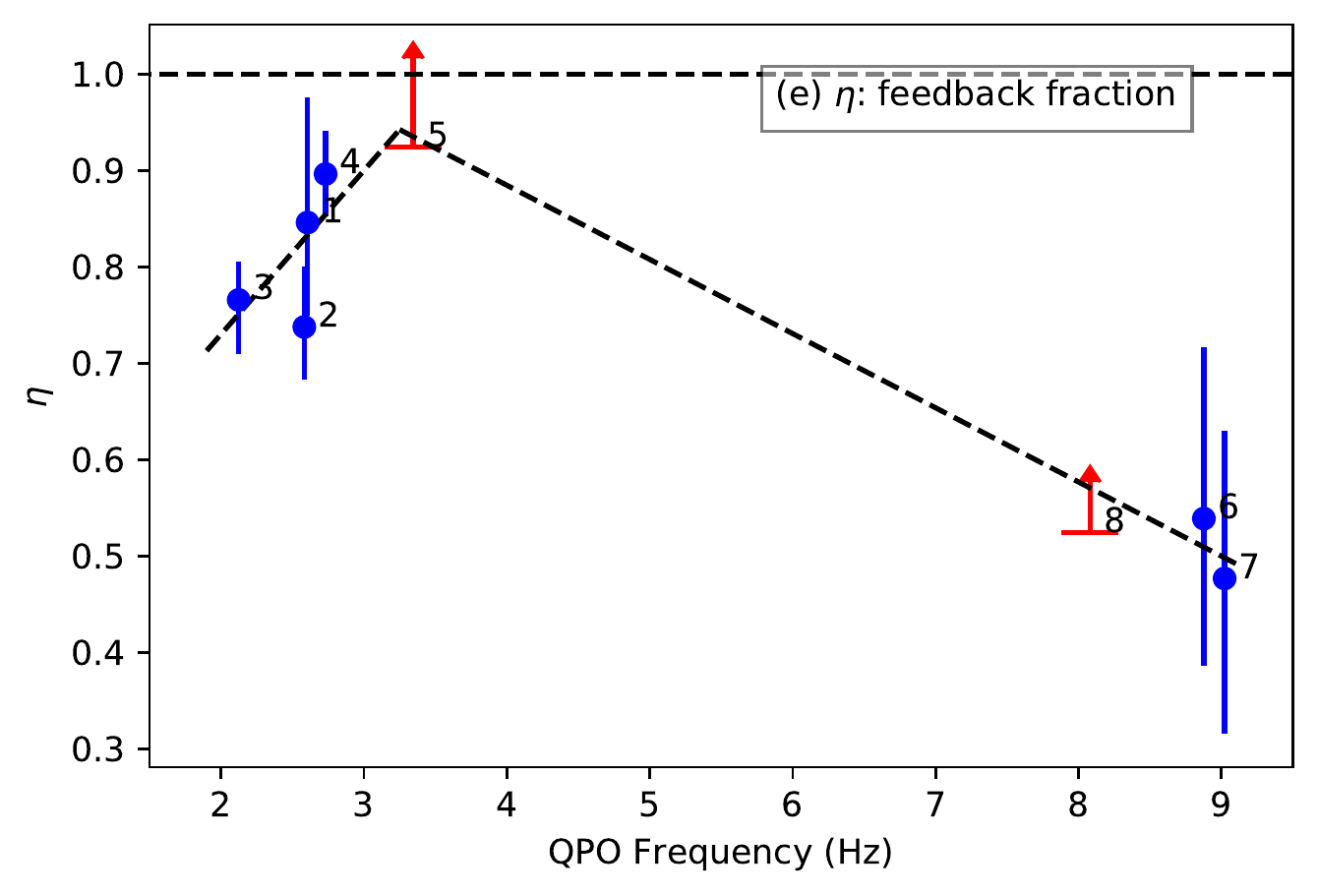}
    \includegraphics[width=0.49\textwidth]{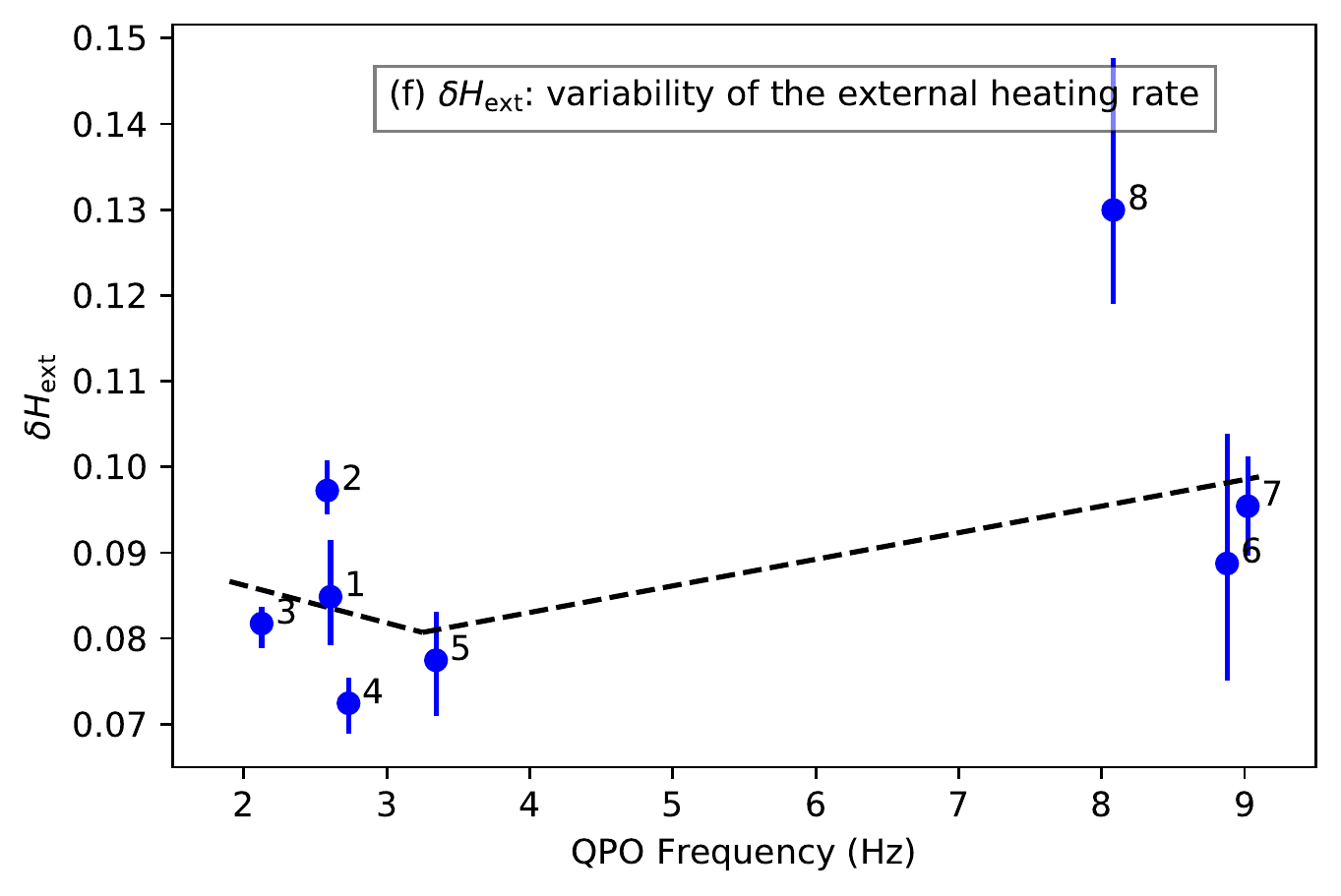}
    \caption{The evolution of the parameters of \maxi vs.\ QPO frequency. The parameters are, respectively, (a) disc temperature, $kT_{\text{s}}$, (b) photon index, $\Gamma$, (c) corona temperature, $kT_{\text{e}}$, (d) corona size, $L$, (e) feedback fraction, $\eta$, and (f) variability of the external heating rate, $\delta\dot{H}_{\text{ext}}$. The error bars represent the 90\% confidence level, while red points show 95\% lower limits. Black dashed broken line is the best fit to the data by a broken line with a break frequency at 3.25~Hz. The numbers on each panel indicate the observations in Table~\ref{tab:qpo}. The horizontal black dashed lines in panels (c) and (e) are the upper boundary of 250~keV and 1, respectively.} 
    \label{fig:evolution}
\end{figure*}

\begin{figure}
    \includegraphics[width=0.49\textwidth]{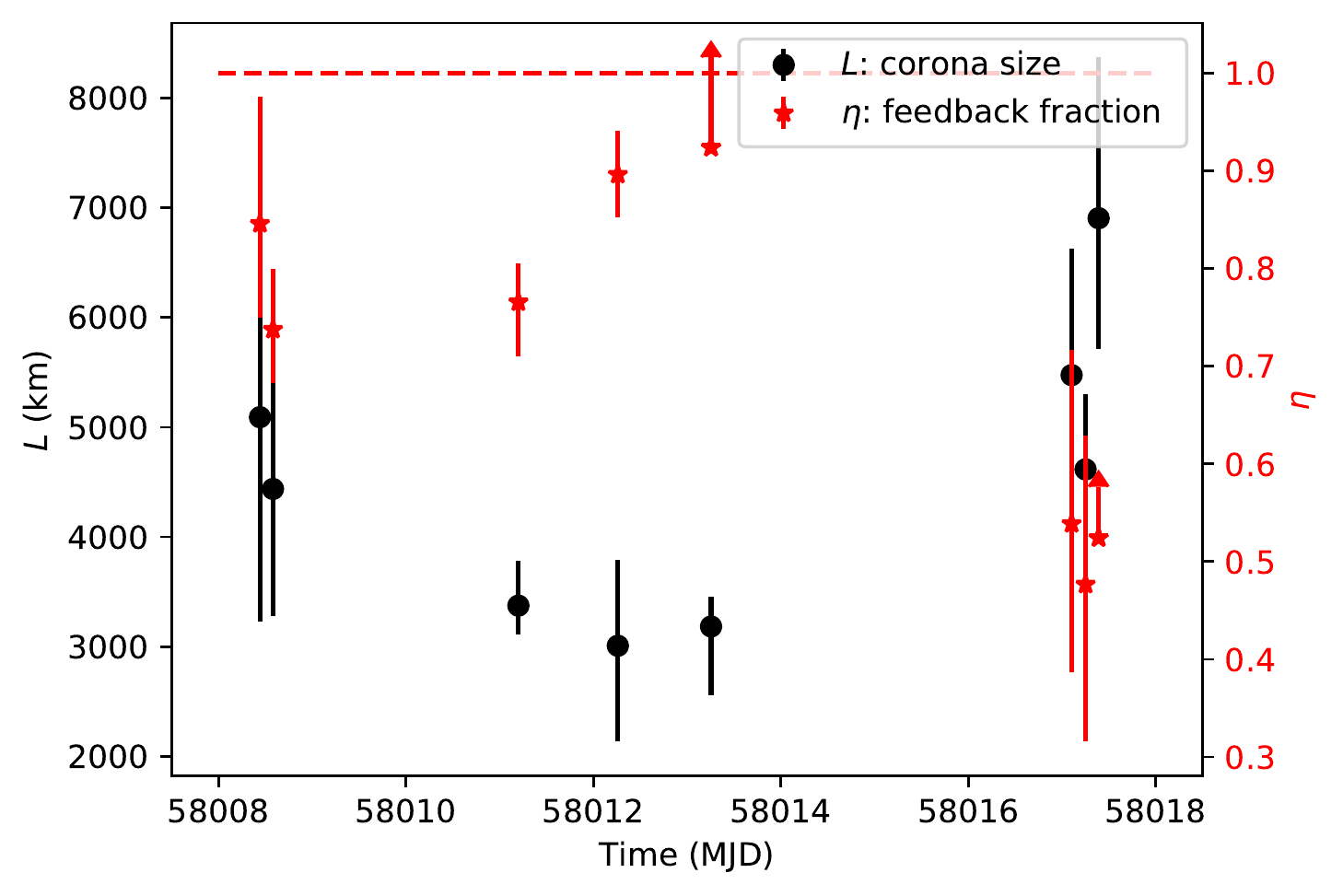}
    \caption{Temporal evolution of the corona size $L$ (black) and the feedback fraction $\eta$ (red) of MAXI~J1535$-$571. The arrow in the point means the lower limit of the measurement. The red horizontal dashed line is the upper limit of 1, according to the definition of the time-dependent Comptonization model.}
    \label{fig:size_and_eta}
\end{figure}

In Figure~\ref{fig:evolution} we show the evolution of the size of the corona, $L$, the feedback fraction, $\eta$, and the amplitude of the variability of the external heating rate, $\delta\dot{H}_{\text{ext}}$, obtained from the time-dependent version of the model \texttt{vkompthdk} that fits the rms and phase-lag spectra. As the frequency of the type-C QPO increases from 2.12~Hz to 3.34~Hz, the size of the corona decreases from 5100~km to 3200~km. At the same time the feedback fraction remains broadly constant in the range of 0.7--1. When the QPO frequency increases further, from 3.34~Hz to 9.02~Hz, the feedback fraction decreases to around 0.5 as the size of the corona increases from 3200~km to around 5500~km. The amplitude of the variability of the external heating rate shows the same evolution trend as the size of the corona. The Lense-Thirring radius~\citep{2009MNRAS.397L.101I}, calculated from the QPO frequency assuming a black hole mass of 10~$M_{\odot}$~\citep{2019MNRAS.487.4221S} and a spin of 0.998, goes from 200~km to 100~km as the QPO frequency evolves from 2~Hz to 9~Hz. These radii are well below the size of the corona, which suggests that the corona always covers a large fraction of the inner disc, consistent with the relatively large values of $\eta$.

If we order the size of the corona and feedback fraction according to time (see Table~\ref{tab:par} and Figure~\ref{fig:size_and_eta}), in observations 1--5 the size of the corona starts at 5100~km, then decreases to 3200~km while, at the same time, the feedback fraction first decreases slightly from 0.85 to 0.74, and then increases marginally and stays at values larger than 0.9. In observations 5--8 the corona size increases from 3200~km to around 5500~km, while the feedback fraction decreases from 0.96 to around 0.5.

Using \textit{NICER} data, Rawat et al. (2022) found that the relation of the time lags of the QPO between two broad energy bands, the spectral parameters of the source, $kT_{\text{s}}$, and $\Gamma$, the corona size $L$ and the feedback fraction $\eta$ vs.\ QPO frequency show a statistically significant break at a QPO frequency $\nu_{\text{c}} = 3.5$~Hz. (Because of \textit{NICER}'s lack of high-energy coverage, different from us they could not measure $kT_{\text{e}}$ in their data.) To explore this in our observations, we first fit each relation in Figure~\ref{fig:evolution} independently with either a straight or a broken line. Since in the latter fits the values of the break frequency, $\nu_{\text{c}}$, of the individual fits are consistent with each other within the 3-$\sigma$ errors, we fit all the relation again simultaneously, this time linking the break frequency of all fits. Differently from Rawat et al. (2022), we find that since the F-test probability is 0.03, the fit with a broken line is only marginally better than that with a straight line, possibly due to the fact that we have no observation with QPO frequency in the range of 4--8~Hz. Our best-fitting broken line gives $\nu_{\text{c}} = 3.25^{+0.60}_{-0.21}$~Hz, consistent with the value found by Rawat el al. (2022). We also fit a broken power-law instead of broken line to the size of the corona and the external heating rate vs. QPO frequency, however, the reduced $\chi^{2}$ and the break frequency do not change significantly compared to the fits with a broken line.

\section{Discussion}\label{sec:discussion}

\begin{figure*}
    \includegraphics[width=0.99\textwidth]{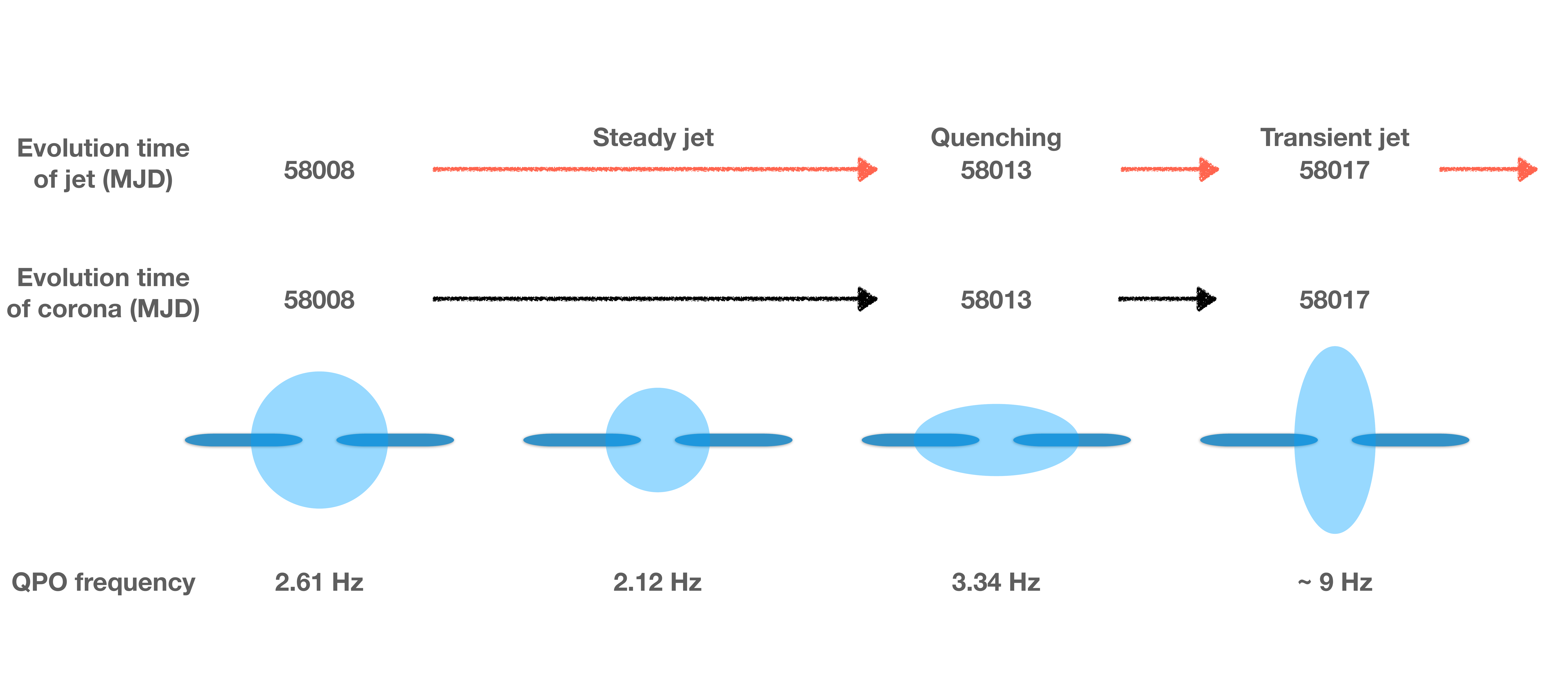}
    \caption{Schematic figure illustrating the time evolution of the corona and the jet of MAXI~J1535$-$571 (see text for a more detailed explanation). From MJD 58008 to MJD 58011 the size of the corona contracts; from MJD 58011 to MJD 58013 the corona contracts vertically but expands horizontally. In the period before MJD 58013 the radio spectrum of the source is consistent with optically thick synchrotron emission from a compact steady jet and on MJD 58013 the jet is quenched. From MJD 58013 to MJD 58017 the corona expands vertically but contracts horizontally and after MJD 58017 the radio data are consistent with optically thin synchrotron emission from a transient jet.} 
    \label{fig:scheme}
\end{figure*}

We have analyzed all the observations of the 2017/2018 outburst of \maxi with \hxmt up to 100~keV. The light curve and `q' path in the HID (Figure~\ref{fig:lc}) show that \maxi behaves like a typical black hole X-ray transient. We find that the temperature of the disc and the photon index of the hard Comptonized component increase with time, indicating a softening of the source spectrum as the source moves along the HIMS. At the same time the inner radius of the accretion disc remains rather stable. We jointly fit the time-averaged, rms and phase-lag spectra of the QPO of eight \hxmt observations of \maxi with type-C QPOs; the frequency of the QPO ranges from 2.12~Hz to 9.02~Hz during the HIMS. As shown in Figure~\ref{fig:evolution}, as the frequency of the type-C QPO increases, the size of the corona, $L$, and the amplitude of the variability of the external heating rate, $\delta\dot{H}_{\text{ext}}$, decrease and then increase slightly. At the same time the feedback fraction, $\eta$, does the opposite.

Previous studies have reported the evolution of the physical parameters of MAXI~J1535$-$571, e.g., spin of the black hole, inclination angle of the accretion disc, and height of the corona. For instance, \citet{2018ApJ...866..122H} analyzed the \hxmt observations, some of which we present in this paper, and concluded that the system inclination angle must be high, judging from the soft lags of the type-C QPOs. From fits to the time-averaged spectra with the reflection model \texttt{rexilllpCp}, \citet{2018ApJ...852L..34X} reported a high spin and a corona that is very close to the black hole in the low hard state. \citet{2018ApJ...860L..28M} used \texttt{relxill} to deduce a high spin and a high emissivity index, which also indicates that the corona is close to the black hole. All of these authors concluded that the inclination angle of \maxi is high. However, the spectral analysis mentioned before just focus on one observation of \textit{NuSTAR} or \textit{NICER}. In the hard/soft intermediate state, we use the \texttt{relxillCp} model to fit eight observations, showing an average spin of 0.998 and inclination angle of $60.2^{\circ}$. Our fits show that the emissivity index is large ($>6$), indicating a corona close to black hole~\citep{2012MNRAS.424.1284W} or a radial dependence of the disc ionization during these observations~\citep{2012A&A...545A.106S}. The time-dependent Comptonization model \texttt{vkompthdk} requires a large corona to explain the lags of the photons. If the corona is large, the radial change of the disc ionization profile could be responsible for the high value of the emissivity index in MAXI~J1535$-$571.

In an outburst of a BHB, as the disc emission increases during the softening of the spectrum in the LHS and HIMS~\citep[e.g.,][]{1997ApJ...479..926M,2018ApJ...866..122H,2020MNRAS.499..851Z}, the rate at which the corona cools down due to inverse-Compton scattering would increase. The temperature of the corona, and hence the energy at which the spectrum cuts off at high energies, may either decrease or increase, depending on the balance between the inverse-Compton cooling and the mechanism that heats up the corona~\citep{2001MNRAS.321..549M,2020MNRAS.492.1399K}. Our results show that during the HIMS, the high-energy cutoff increases from 32 keV to at least 250 keV. This trend is the same as the ones shown in~\citet{2008ApJ...679..655J}, \citet{2008MNRAS.390..227D} and \citet{2009MNRAS.400.1603M}. \citet{2008MNRAS.390..227D} showed this phenomenology in the HIMS data of GRO~J1655$-$40 with INTEGRAL, which displays a non-thermal hard tail at high energy. This tail becomes more dominant and the high-energy cutoff increases when the source is about to transit into the soft state. Similarly, \citet{2009MNRAS.400.1603M} showed that in the case of GX~339$-$4 the power-law cutoff energy first decreases as the source moves in the LHS and HIMS, and increases again just before the transition to the SIMS. The evolution of the power-law cutoff energy in these two sources is consistent with the behavior of the corona temperature that we report here for \maxi in the HIMS.

In the framework of Lense-Thirring (LT) precession as the explanation of the dynamical origin of the type-C QPOs, \citet{2009MNRAS.397L.101I} assumed that a torus-like corona precesses within the inner radius of a truncated disc~\citep[see also,][]{2018ApJ...858...82Y}. Assuming that the mass of the black hole in \maxi is $10 M_{\odot}$~\citep{2019MNRAS.487.4221S} and the spin is 0.998 from our fits, the LT radius decreases from 200~km to 100~km as the QPO frequency increases from 2~Hz to 9~Hz. The LT radius is much smaller than the corona size we obtain from our fits, the minimum of which is around 3000~km. The time-dependent Comptonization model~\citep[][]{2020MNRAS.492.1399K} that we use does not explain the dynamical origin of the QPO, but only requires that there is a coupled oscillation of the physical quantities: corona temperature, $kT_{\text{e}}$, temperature of the inner accretion disc, $kT_{\text{s}}$, external heating rate, $\delta\dot{H}_{\text{ext}}$, and the time-dependent photon number density that is dynamically scattering in the plasma. We favor the scenario in which the inner disc oscillates since, if the source of soft photons precesses, the line-of-sight soft photons oscillate and would potentially cause a perturbation in the corona temperature.

Assuming that the time lags of the broadband noise component are due to reverberation, \citet{2019Natur.565..198K} concluded that the corona of MAXI~J1820+070 contracts as the source is in the low-hard state. \citet{2021ApJ...910L...3W} proposed that the source expands again when evolving further in the intermediate state. Using the time-dependent Comptonization model, \citet{2021MNRAS.503.5522K} showed that in \grs the size of the corona first decreases from 10000~km to 100~km and then increases to 1000~km as the QPO frequency increases from 0.5~Hz to 6~Hz, while the feedback fraction increases monotonically~\citep[see also][]{mendez}. Our result of the size of the corona shows the same trend in \maxi as that in GRS~1915+105, but the size of the corona in the case of \maxi is about one order of magnitude larger than in GRS~1915+105. In MAXI~J1535$-$571, different from \grs the feedback fraction first remains broadly constant in the range of 0.7--1 and then decreases as the QPO frequency increases, providing a different picture of the evolution of the corona.

Similar to the case of \grs\citep{mendez}, the large size of the corona in \maxi indicates that the corona would cover the disc beyond the inner radius. However, at the same time that the corona is large, the feedback fraction is small. This suggests that the corona is not spherical, but extends perpendicularly to the accretion disc. In these conditions the assumption of a spherical corona in the model of~\citet{2020MNRAS.492.1399K} would not be satisfied. In this case $L$ must therefore represent a characteristic size of the corona such that the magnitude of the predicted lags can match the observed ones~\citep[see][]{mendez}.

Our results of the temporal evolution of the corona may be compared to the evolution of the radio jet detected in this source. Radio measurements~\citep{2019ApJ...883..198R,2020MNRAS.498.5772R,2021PASA...38...45C} show that during the intermediate state of \maxi a steady, optically thick jet gradually quenches and then changes into a transient jet. More specifically, on MJD 58008, the radio observations detect a steady, optically thick jet. On MJD 58013 the radio emission weakens and the jet is quenched. On MJD 58017, the radio emission reappears, but this time the spectrum is optically thin, indicating a transient jet with discrete ejecta.

Based on our results and the radio measurements of~\citet{2019ApJ...883..198R} and~\citet{2021PASA...38...45C}, in Figure~\ref{fig:scheme}, we present schematically a possible scenario of the evolution of the X-ray corona and the simultaneous radio jet. The parameters from the reflection suggest a rather stable inner radius of the disc that is very close to the ISCO, so the disc inner radius does not change significantly in Figure~\ref{fig:scheme}. Initially, the QPO frequency is 2.61~Hz, and the size of the corona is 5100~km. When the QPO frequency first decreases from 2.61~Hz to 2.12~Hz, the corona size decreases from 5100~km to 3400~km and the feedback fraction decreases slightly from 0.85 to 0.77 in the time period from MJD 58008 to MJD 58011 (Figure~\ref{fig:size_and_eta}). As the QPO frequency increases from 2.12~Hz to 3.34~Hz, the corona size remains constant at around 3000--3400~km. Since the feedback fraction increases from 0.77 to 0.96, suggesting that the corona covers a larger part of the disc, the corona should expand along the horizontal direction but contract along the vertical direction. This phase corresponds to the time period from MJD 58011 to MJD 58013. In the period MJD 58013--58017, the QPO changes from 3.34~Hz to 8.08~Hz and the corona expands again. Since in this phase the feedback fraction decreases from 0.96 to $\sim$ 0.5, indicating that a smaller part of the disc is covered by the corona, the expansion of the corona must be along the vertical direction.

Using different approaches, a similar evolution of the corona has been reported. Using the reflection model \texttt{relxillCp}~\citep{2014ApJ...782...76G} to fit the \hxmt data, \citet{2021NatCo..12.1025Y} found that the corona in MAXI~J1820+070 outflows faster when it moves closer to the black hole, which suggests a jet-like corona and that the jet gains energy as the corona outflows. Utilizing the \texttt{reltrans} model~\citep{2019MNRAS.488..324I} to fit the broadband time-lag spectra, \citet{2021ApJ...910L...3W} found a connection between the corona and the jet in the intermediate state of MAXI~J1820+070. The picture proposed by~\citet{2021ApJ...910L...3W} is in agreement with ours: The quenching of the steady jet follows the contraction of the corona, while the transient jet appears after the expansion of the corona. More observations and spectral modelling using the time-dependent Comptonization model in the future would provide a complete picture of the corona-jet evolution, and the role of the disc during this evolution.

\nocite{bellavita,rawat,mendez}


\section*{Acknowledgements}

We would like to thank the referee for constructive comments that helped improve this paper. This work is supported by the National Key R\&D Program of China (2021YFA0718500). We thank Professor Bei You for helpful discussions. Y.Z.\ acknowledges support from China Scholarship Council (CSC 201906100030). M.M.\ and F.G.\ acknowledge the research programme Athena with project number 184.034.002, which is (partly) financed by the Dutch Research Council (NWO). F.G.\ is a CONICET researcher. F.G.\ acknowledges support by PIP 0113 (CONICET). This work received financial support from PICT-2017-2865 (ANPCyT). T.M.B.\ acknowledges financial contribution from the agreement ASI-INAF n.\ 2017-14-H.0 and from PRIN INAF 2019 n.15. S.Z.\ acknowledges support of the National Natural Science Foundation of China under grant U1838201. L.T.\ acknowledges funding support from the National Natural Science Foundation of China (NSFC) under grant No.\ 12122306. D.A.\ acknowledges support from the Royal Society. L.Z.\ acknowledges support from the Royal Society Newton Funds. C.B.\ is a CIN fellow.


\section*{Data Availability}

The data used in this article are accessible at the \hxmt website \url{http://hxmtweb.ihep.ac.cn}.
 



\bibliographystyle{mnras}
\bibliography{reference} 

\begin{thebibliography}{}
\makeatletter
\relax
\def\mn@urlcharsother{\let\do\@makeother \do\$\do\&\do\#\do\^\do\_\do\%\do\~}
\def\mn@doi{\begingroup\mn@urlcharsother \@ifnextchar [ {\mn@doi@}
  {\mn@doi@[]}}
\def\mn@doi@[#1]#2{\def\@tempa{#1}\ifx\@tempa\@empty \href
  {http://dx.doi.org/#2} {doi:#2}\else \href {http://dx.doi.org/#2} {#1}\fi
  \endgroup}
\def\mn@eprint#1#2{\mn@eprint@#1:#2::\@nil}
\def\mn@eprint@arXiv#1{\href {http://arxiv.org/abs/#1} {{\tt arXiv:#1}}}
\def\mn@eprint@dblp#1{\href {http://dblp.uni-trier.de/rec/bibtex/#1.xml}
  {dblp:#1}}
\def\mn@eprint@#1:#2:#3:#4\@nil{\def\@tempa {#1}\def\@tempb {#2}\def\@tempc
  {#3}\ifx \@tempc \@empty \let \@tempc \@tempb \let \@tempb \@tempa \fi \ifx
  \@tempb \@empty \def\@tempb {arXiv}\fi \@ifundefined
  {mn@eprint@\@tempb}{\@tempb:\@tempc}{\expandafter \expandafter \csname
  mn@eprint@\@tempb\endcsname \expandafter{\@tempc}}}

\bibitem[\protect\citeauthoryear{{Arnaud}}{{Arnaud}}{1996}]{1996ASPC..101...17A}
{Arnaud} K.~A.,  1996, in {Jacoby} G.~H.,  {Barnes} J.,  eds,  Astronomical
  Society of the Pacific Conference Series Vol. 101, Astronomical Data Analysis
  Software and Systems V. p.~17

\bibitem[\protect\citeauthoryear{{Bambi} et~al.,}{{Bambi}
  et~al.}{2021}]{2021SSRv..217...65B}
{Bambi} C.,  et~al., 2021, \mn@doi [\ssr] {10.1007/s11214-021-00841-8}, \href
  {https://ui.adsabs.harvard.edu/abs/2021SSRv..217...65B} {217, 65}

\bibitem[\protect\citeauthoryear{{Bellavita} \& {et al.,}}{{Bellavita} \& {et
  al.,}}{2022}]{bellavita}
{Bellavita} C.,  {et al.,} 2022, in prep

\bibitem[\protect\citeauthoryear{{Belloni} \& {Hasinger}}{{Belloni} \&
  {Hasinger}}{1990}]{1990A&A...230..103B}
{Belloni} T.,  {Hasinger} G.,  1990, \aap, \href
  {https://ui.adsabs.harvard.edu/abs/1990A&A...230..103B} {230, 103}

\bibitem[\protect\citeauthoryear{{Belloni}, {Psaltis}  \& {van der
  Klis}}{{Belloni} et~al.}{2002}]{2002ApJ...572..392B}
{Belloni} T.,  {Psaltis} D.,   {van der Klis} M.,  2002, \mn@doi [\apj]
  {10.1086/340290}, \href
  {https://ui.adsabs.harvard.edu/abs/2002ApJ...572..392B} {572, 392}

\bibitem[\protect\citeauthoryear{{Belloni}, {Homan}, {Casella}, {van der Klis},
  {Nespoli}, {Lewin}, {Miller}  \& {M{\'e}ndez}}{{Belloni}
  et~al.}{2005}]{2005A&A...440..207B}
{Belloni} T.,  {Homan} J.,  {Casella} P.,  {van der Klis} M.,  {Nespoli} E.,
  {Lewin} W.~H.~G.,  {Miller} J.~M.,   {M{\'e}ndez} M.,  2005, \mn@doi [\aap]
  {10.1051/0004-6361:20042457}, \href
  {https://ui.adsabs.harvard.edu/abs/2005A&A...440..207B} {440, 207}

\bibitem[\protect\citeauthoryear{{Bhargava}, {Belloni}, {Bhattacharya}  \&
  {Misra}}{{Bhargava} et~al.}{2019}]{2019MNRAS.488..720B}
{Bhargava} Y.,  {Belloni} T.,  {Bhattacharya} D.,   {Misra} R.,  2019, \mn@doi
  [\mnras] {10.1093/mnras/stz1774}, \href
  {https://ui.adsabs.harvard.edu/abs/2019MNRAS.488..720B} {488, 720}

\bibitem[\protect\citeauthoryear{{Bu} et~al.,}{{Bu}
  et~al.}{2021}]{2021ApJ...919...92B}
{Bu} Q.~C.,  et~al., 2021, \mn@doi [\apj] {10.3847/1538-4357/ac11f5}, \href
  {https://ui.adsabs.harvard.edu/abs/2021ApJ...919...92B} {919, 92}

\bibitem[\protect\citeauthoryear{{Casella}, {Belloni}  \& {Stella}}{{Casella}
  et~al.}{2005}]{2005ApJ...629..403C}
{Casella} P.,  {Belloni} T.,   {Stella} L.,  2005, \mn@doi [\apj]
  {10.1086/431174}, \href
  {https://ui.adsabs.harvard.edu/abs/2005ApJ...629..403C} {629, 403}

\bibitem[\protect\citeauthoryear{{Chauhan} et~al.,}{{Chauhan}
  et~al.}{2021}]{2021PASA...38...45C}
{Chauhan} J.,  et~al., 2021, \mn@doi [\pasa] {10.1017/pasa.2021.38}, \href
  {https://ui.adsabs.harvard.edu/abs/2021PASA...38...45C} {38, e045}

\bibitem[\protect\citeauthoryear{{Corbel}, {Fender}, {Tomsick}, {Tzioumis}  \&
  {Tingay}}{{Corbel} et~al.}{2004}]{2004ApJ...617.1272C}
{Corbel} S.,  {Fender} R.~P.,  {Tomsick} J.~A.,  {Tzioumis} A.~K.,   {Tingay}
  S.,  2004, \mn@doi [\apj] {10.1086/425650}, \href
  {https://ui.adsabs.harvard.edu/abs/2004ApJ...617.1272C} {617, 1272}

\bibitem[\protect\citeauthoryear{{Del Santo}, {Malzac}, {Jourdain}, {Belloni}
  \& {Ubertini}}{{Del Santo} et~al.}{2008}]{2008MNRAS.390..227D}
{Del Santo} M.,  {Malzac} J.,  {Jourdain} E.,  {Belloni} T.,   {Ubertini} P.,
  2008, \mn@doi [\mnras] {10.1111/j.1365-2966.2008.13672.x}, \href
  {https://ui.adsabs.harvard.edu/abs/2008MNRAS.390..227D} {390, 227}

\bibitem[\protect\citeauthoryear{{Done}, {Gierli{\'n}ski}  \& {Kubota}}{{Done}
  et~al.}{2007}]{2007A&ARv..15....1D}
{Done} C.,  {Gierli{\'n}ski} M.,   {Kubota} A.,  2007, \mn@doi [\aapr]
  {10.1007/s00159-007-0006-1}, \href
  {https://ui.adsabs.harvard.edu/abs/2007A&ARv..15....1D} {15, 1}

\bibitem[\protect\citeauthoryear{{Fabian}, {Rees}, {Stella}  \&
  {White}}{{Fabian} et~al.}{1989}]{1989MNRAS.238..729F}
{Fabian} A.~C.,  {Rees} M.~J.,  {Stella} L.,   {White} N.~E.,  1989, \mn@doi
  [\mnras] {10.1093/mnras/238.3.729}, \href
  {https://ui.adsabs.harvard.edu/abs/1989MNRAS.238..729F} {238, 729}

\bibitem[\protect\citeauthoryear{{Fender}}{{Fender}}{2001}]{2001MNRAS.322...31F}
{Fender} R.~P.,  2001, \mn@doi [\mnras] {10.1046/j.1365-8711.2001.04080.x},
  \href {https://ui.adsabs.harvard.edu/abs/2001MNRAS.322...31F} {322, 31}

\bibitem[\protect\citeauthoryear{{Fender}}{{Fender}}{2006}]{2006csxs.book..381F}
{Fender} R.,  2006, {Jets from X-ray binaries}.
pp 381--419

\bibitem[\protect\citeauthoryear{{Fender}, {Belloni}  \& {Gallo}}{{Fender}
  et~al.}{2004}]{2004MNRAS.355.1105F}
{Fender} R.~P.,  {Belloni} T.~M.,   {Gallo} E.,  2004, \mn@doi [\mnras]
  {10.1111/j.1365-2966.2004.08384.x}, \href
  {https://ui.adsabs.harvard.edu/abs/2004MNRAS.355.1105F} {355, 1105}

\bibitem[\protect\citeauthoryear{{Galeev}, {Rosner}  \& {Vaiana}}{{Galeev}
  et~al.}{1979}]{1979ApJ...229..318G}
{Galeev} A.~A.,  {Rosner} R.,   {Vaiana} G.~S.,  1979, \mn@doi [\apj]
  {10.1086/156957}, \href
  {https://ui.adsabs.harvard.edu/abs/1979ApJ...229..318G} {229, 318}

\bibitem[\protect\citeauthoryear{{Garc{\'\i}a} et~al.,}{{Garc{\'\i}a}
  et~al.}{2014}]{2014ApJ...782...76G}
{Garc{\'\i}a} J.,  et~al., 2014, \mn@doi [\apj] {10.1088/0004-637X/782/2/76},
  \href {https://ui.adsabs.harvard.edu/abs/2014ApJ...782...76G} {782, 76}

\bibitem[\protect\citeauthoryear{{Garc{\'\i}a} et~al.,}{{Garc{\'\i}a}
  et~al.}{2018}]{2018ApJ...864...25G}
{Garc{\'\i}a} J.~A.,  et~al., 2018, \mn@doi [\apj] {10.3847/1538-4357/aad231},
  \href {https://ui.adsabs.harvard.edu/abs/2018ApJ...864...25G} {864, 25}

\bibitem[\protect\citeauthoryear{{Garc{\'\i}a}, {M{\'e}ndez}, {Karpouzas},
  {Belloni}, {Zhang}  \& {Altamirano}}{{Garc{\'\i}a}
  et~al.}{2021}]{2021MNRAS.501.3173G}
{Garc{\'\i}a} F.,  {M{\'e}ndez} M.,  {Karpouzas} K.,  {Belloni} T.,  {Zhang}
  L.,   {Altamirano} D.,  2021, \mn@doi [\mnras] {10.1093/mnras/staa3944},
  \href {https://ui.adsabs.harvard.edu/abs/2021MNRAS.501.3173G} {501, 3173}

\bibitem[\protect\citeauthoryear{{Gilfanov}}{{Gilfanov}}{2010}]{2010LNP...794...17G}
{Gilfanov} M.,  2010, {X-Ray Emission from Black-Hole Binaries}.
p.~17, \mn@doi{10.1007/978-3-540-76937-8\_2}

\bibitem[\protect\citeauthoryear{{Haardt} \& {Maraschi}}{{Haardt} \&
  {Maraschi}}{1991}]{1991ApJ...380L..51H}
{Haardt} F.,  {Maraschi} L.,  1991, \mn@doi [\apjl] {10.1086/186171}, \href
  {https://ui.adsabs.harvard.edu/abs/1991ApJ...380L..51H} {380, L51}

\bibitem[\protect\citeauthoryear{{Homan}, {Wijnands}, {van der Klis},
  {Belloni}, {van Paradijs}, {Klein-Wolt}, {Fender}  \& {M{\'e}ndez}}{{Homan}
  et~al.}{2001}]{2001ApJS..132..377H}
{Homan} J.,  {Wijnands} R.,  {van der Klis} M.,  {Belloni} T.,  {van Paradijs}
  J.,  {Klein-Wolt} M.,  {Fender} R.,   {M{\'e}ndez} M.,  2001, \mn@doi [\apjs]
  {10.1086/318954}, \href
  {https://ui.adsabs.harvard.edu/abs/2001ApJS..132..377H} {132, 377}

\bibitem[\protect\citeauthoryear{{Huang} et~al.,}{{Huang}
  et~al.}{2018}]{2018ApJ...866..122H}
{Huang} Y.,  et~al., 2018, \mn@doi [\apj] {10.3847/1538-4357/aade4c}, \href
  {https://ui.adsabs.harvard.edu/abs/2018ApJ...866..122H} {866, 122}

\bibitem[\protect\citeauthoryear{{Ingram} \& {Motta}}{{Ingram} \&
  {Motta}}{2019}]{2019NewAR..8501524I}
{Ingram} A.~R.,  {Motta} S.~E.,  2019, \mn@doi [\nar]
  {10.1016/j.newar.2020.101524}, \href
  {https://ui.adsabs.harvard.edu/abs/2019NewAR..8501524I} {85, 101524}

\bibitem[\protect\citeauthoryear{{Ingram}, {Done}  \& {Fragile}}{{Ingram}
  et~al.}{2009}]{2009MNRAS.397L.101I}
{Ingram} A.,  {Done} C.,   {Fragile} P.~C.,  2009, \mn@doi [\mnras]
  {10.1111/j.1745-3933.2009.00693.x}, \href
  {https://ui.adsabs.harvard.edu/abs/2009MNRAS.397L.101I} {397, L101}

\bibitem[\protect\citeauthoryear{{Ingram}, {Mastroserio}, {Dauser},
  {Hovenkamp}, {van der Klis}  \& {Garc{\'\i}a}}{{Ingram}
  et~al.}{2019}]{2019MNRAS.488..324I}
{Ingram} A.,  {Mastroserio} G.,  {Dauser} T.,  {Hovenkamp} P.,  {van der Klis}
  M.,   {Garc{\'\i}a} J.~A.,  2019, \mn@doi [\mnras] {10.1093/mnras/stz1720},
  \href {https://ui.adsabs.harvard.edu/abs/2019MNRAS.488..324I} {488, 324}

\bibitem[\protect\citeauthoryear{{Jiang}, {F{\"u}rst}, {Walton}, {Parker}  \&
  {Fabian}}{{Jiang} et~al.}{2020}]{2020MNRAS.492.1947J}
{Jiang} J.,  {F{\"u}rst} F.,  {Walton} D.~J.,  {Parker} M.~L.,   {Fabian}
  A.~C.,  2020, \mn@doi [\mnras] {10.1093/mnras/staa017}, \href
  {https://ui.adsabs.harvard.edu/abs/2020MNRAS.492.1947J} {492, 1947}

\bibitem[\protect\citeauthoryear{{Joinet}, {Kalemci}  \& {Senziani}}{{Joinet}
  et~al.}{2008}]{2008ApJ...679..655J}
{Joinet} A.,  {Kalemci} E.,   {Senziani} F.,  2008, \mn@doi [\apj]
  {10.1086/533512}, \href
  {https://ui.adsabs.harvard.edu/abs/2008ApJ...679..655J} {679, 655}

\bibitem[\protect\citeauthoryear{{Kara} et~al.,}{{Kara}
  et~al.}{2019}]{2019Natur.565..198K}
{Kara} E.,  et~al., 2019, \mn@doi [\nat] {10.1038/s41586-018-0803-x}, \href
  {https://ui.adsabs.harvard.edu/abs/2019Natur.565..198K} {565, 198}

\bibitem[\protect\citeauthoryear{{Karpouzas}, {M{\'e}ndez}, {Ribeiro},
  {Altamirano}, {Blaes}  \& {Garc{\'\i}a}}{{Karpouzas}
  et~al.}{2020}]{2020MNRAS.492.1399K}
{Karpouzas} K.,  {M{\'e}ndez} M.,  {Ribeiro} E.~M.,  {Altamirano} D.,  {Blaes}
  O.,   {Garc{\'\i}a} F.,  2020, \mn@doi [\mnras] {10.1093/mnras/stz3502},
  \href {https://ui.adsabs.harvard.edu/abs/2020MNRAS.492.1399K} {492, 1399}

\bibitem[\protect\citeauthoryear{{Karpouzas}, {M{\'e}ndez}, {Garc{\'\i}a},
  {Zhang}, {Altamirano}, {Belloni}  \& {Zhang}}{{Karpouzas}
  et~al.}{2021}]{2021MNRAS.503.5522K}
{Karpouzas} K.,  {M{\'e}ndez} M.,  {Garc{\'\i}a} F.,  {Zhang} L.,  {Altamirano}
  D.,  {Belloni} T.,   {Zhang} Y.,  2021, \mn@doi [\mnras]
  {10.1093/mnras/stab827}, \href
  {https://ui.adsabs.harvard.edu/abs/2021MNRAS.503.5522K} {503, 5522}

\bibitem[\protect\citeauthoryear{{Kennea}, {Evans}, {Beardmore}, {Krimm},
  {Romano}, {Yamaoka}, {Serino}  \& {Negoro}}{{Kennea}
  et~al.}{2017}]{2017ATel10700....1K}
{Kennea} J.~A.,  {Evans} P.~A.,  {Beardmore} A.~P.,  {Krimm} H.~A.,  {Romano}
  P.,  {Yamaoka} K.,  {Serino} M.,   {Negoro} H.,  2017, The Astronomer's
  Telegram, \href {https://ui.adsabs.harvard.edu/abs/2017ATel10700....1K}
  {10700, 1}

\bibitem[\protect\citeauthoryear{{Kong} et~al.,}{{Kong}
  et~al.}{2020}]{2020JHEAp..25...29K}
{Kong} L.~D.,  et~al., 2020, \mn@doi [Journal of High Energy Astrophysics]
  {10.1016/j.jheap.2020.01.003}, \href
  {https://ui.adsabs.harvard.edu/abs/2020JHEAp..25...29K} {25, 29}

\bibitem[\protect\citeauthoryear{{Kumar} \& {Misra}}{{Kumar} \&
  {Misra}}{2014}]{2014MNRAS.445.2818K}
{Kumar} N.,  {Misra} R.,  2014, \mn@doi [\mnras] {10.1093/mnras/stu1946}, \href
  {https://ui.adsabs.harvard.edu/abs/2014MNRAS.445.2818K} {445, 2818}

\bibitem[\protect\citeauthoryear{{Lee} \& {Miller}}{{Lee} \&
  {Miller}}{1998}]{1998MNRAS.299..479L}
{Lee} H.~C.,  {Miller} G.~S.,  1998, \mn@doi [\mnras]
  {10.1046/j.1365-8711.1998.01842.x}, \href
  {https://ui.adsabs.harvard.edu/abs/1998MNRAS.299..479L} {299, 479}

\bibitem[\protect\citeauthoryear{{Lee}, {Misra}  \& {Taam}}{{Lee}
  et~al.}{2001}]{2001ApJ...549L.229L}
{Lee} H.~C.,  {Misra} R.,   {Taam} R.~E.,  2001, \mn@doi [\apjl]
  {10.1086/319171}, \href
  {https://ui.adsabs.harvard.edu/abs/2001ApJ...549L.229L} {549, L229}

\bibitem[\protect\citeauthoryear{{Ma} et~al.,}{{Ma}
  et~al.}{2021}]{2021NatAs...5...94M}
{Ma} X.,  et~al., 2021, \mn@doi [Nature Astronomy] {10.1038/s41550-020-1192-2},
  \href {https://ui.adsabs.harvard.edu/abs/2021NatAs...5...94M} {5, 94}

\bibitem[\protect\citeauthoryear{{Maccarone} \& {Coppi}}{{Maccarone} \&
  {Coppi}}{2003}]{2003MNRAS.338..189M}
{Maccarone} T.~J.,  {Coppi} P.~S.,  2003, \mn@doi [\mnras]
  {10.1046/j.1365-8711.2003.06040.x}, \href
  {https://ui.adsabs.harvard.edu/abs/2003MNRAS.338..189M} {338, 189}

\bibitem[\protect\citeauthoryear{{M{\'e}ndez} \& {van der Klis}}{{M{\'e}ndez}
  \& {van der Klis}}{1997}]{1997ApJ...479..926M}
{M{\'e}ndez} M.,  {van der Klis} M.,  1997, \mn@doi [\apj] {10.1086/303914},
  \href {https://ui.adsabs.harvard.edu/abs/1997ApJ...479..926M} {479, 926}

\bibitem[\protect\citeauthoryear{{M{\'e}ndez}, {Altamirano}, {Belloni}  \&
  {Sanna}}{{M{\'e}ndez} et~al.}{2013}]{2013MNRAS.435.2132M}
{M{\'e}ndez} M.,  {Altamirano} D.,  {Belloni} T.,   {Sanna} A.,  2013, \mn@doi
  [\mnras] {10.1093/mnras/stt1431}, \href
  {https://ui.adsabs.harvard.edu/abs/2013MNRAS.435.2132M} {435, 2132}

\bibitem[\protect\citeauthoryear{{M{\'e}ndez}, {Karpouzas}, {Garc{\'\i}a},
  {Zhang}, {Zhang}, {Belloni}  \& {Altamirano}}{{M{\'e}ndez}
  et~al.}{2022}]{mendez}
{M{\'e}ndez} M.,  {Karpouzas} K.,  {Garc{\'\i}a} F.,  {Zhang} L.,  {Zhang} Y.,
  {Belloni} T.,   {Altamirano} D.,  2022, Nature Astronomy, in press

\bibitem[\protect\citeauthoryear{{Merloni} \& {Fabian}}{{Merloni} \&
  {Fabian}}{2001}]{2001MNRAS.321..549M}
{Merloni} A.,  {Fabian} A.~C.,  2001, \mn@doi [\mnras]
  {10.1046/j.1365-8711.2001.04060.x}, \href
  {https://ui.adsabs.harvard.edu/abs/2001MNRAS.321..549M} {321, 549}

\bibitem[\protect\citeauthoryear{{Miller-Jones} et~al.,}{{Miller-Jones}
  et~al.}{2012}]{2012MNRAS.421..468M}
{Miller-Jones} J.~C.~A.,  et~al., 2012, \mn@doi [\mnras]
  {10.1111/j.1365-2966.2011.20326.x}, \href
  {https://ui.adsabs.harvard.edu/abs/2012MNRAS.421..468M} {421, 468}

\bibitem[\protect\citeauthoryear{{Miller} et~al.,}{{Miller}
  et~al.}{2018}]{2018ApJ...860L..28M}
{Miller} J.~M.,  et~al., 2018, \mn@doi [\apjl] {10.3847/2041-8213/aacc61},
  \href {https://ui.adsabs.harvard.edu/abs/2018ApJ...860L..28M} {860, L28}

\bibitem[\protect\citeauthoryear{{Mirabel} \& {Rodr{\'\i}guez}}{{Mirabel} \&
  {Rodr{\'\i}guez}}{1994}]{1994Natur.371...46M}
{Mirabel} I.~F.,  {Rodr{\'\i}guez} L.~F.,  1994, \mn@doi [\nat]
  {10.1038/371046a0}, \href
  {https://ui.adsabs.harvard.edu/abs/1994Natur.371...46M} {371, 46}

\bibitem[\protect\citeauthoryear{{Mitsuda} et~al.,}{{Mitsuda}
  et~al.}{1984}]{1984PASJ...36..741M}
{Mitsuda} K.,  et~al., 1984, \pasj, \href
  {https://ui.adsabs.harvard.edu/abs/1984PASJ...36..741M} {36, 741}

\bibitem[\protect\citeauthoryear{{Miyamoto}, {Kitamoto}, {Mitsuda}  \&
  {Dotani}}{{Miyamoto} et~al.}{1988}]{1988Natur.336..450M}
{Miyamoto} S.,  {Kitamoto} S.,  {Mitsuda} K.,   {Dotani} T.,  1988, \mn@doi
  [\nat] {10.1038/336450a0}, \href
  {https://ui.adsabs.harvard.edu/abs/1988Natur.336..450M} {336, 450}

\bibitem[\protect\citeauthoryear{{Morgan}, {Remillard}  \& {Greiner}}{{Morgan}
  et~al.}{1997}]{1997ApJ...482..993M}
{Morgan} E.~H.,  {Remillard} R.~A.,   {Greiner} J.,  1997, \mn@doi [\apj]
  {10.1086/304191}, \href
  {https://ui.adsabs.harvard.edu/abs/1997ApJ...482..993M} {482, 993}

\bibitem[\protect\citeauthoryear{{Motta}, {Belloni}  \& {Homan}}{{Motta}
  et~al.}{2009}]{2009MNRAS.400.1603M}
{Motta} S.,  {Belloni} T.,   {Homan} J.,  2009, \mn@doi [\mnras]
  {10.1111/j.1365-2966.2009.15566.x}, \href
  {https://ui.adsabs.harvard.edu/abs/2009MNRAS.400.1603M} {400, 1603}

\bibitem[\protect\citeauthoryear{{Mu{\~n}oz-Darias}, {Motta}  \&
  {Belloni}}{{Mu{\~n}oz-Darias} et~al.}{2011}]{2011MNRAS.410..679M}
{Mu{\~n}oz-Darias} T.,  {Motta} S.,   {Belloni} T.~M.,  2011, \mn@doi [\mnras]
  {10.1111/j.1365-2966.2010.17476.x}, \href
  {https://ui.adsabs.harvard.edu/abs/2011MNRAS.410..679M} {410, 679}

\bibitem[\protect\citeauthoryear{{Negoro} et~al.,}{{Negoro}
  et~al.}{2017}]{2017ATel10708....1N}
{Negoro} H.,  et~al., 2017, The Astronomer's Telegram, \href
  {https://ui.adsabs.harvard.edu/abs/2017ATel10708....1N} {10708, 1}

\bibitem[\protect\citeauthoryear{{Nowak}}{{Nowak}}{2000}]{2000MNRAS.318..361N}
{Nowak} M.~A.,  2000, \mn@doi [\mnras] {10.1046/j.1365-8711.2000.03668.x},
  \href {https://ui.adsabs.harvard.edu/abs/2000MNRAS.318..361N} {318, 361}

\bibitem[\protect\citeauthoryear{{Nowak}, {Vaughan}, {Wilms}, {Dove}  \&
  {Begelman}}{{Nowak} et~al.}{1999}]{1999ApJ...510..874N}
{Nowak} M.~A.,  {Vaughan} B.~A.,  {Wilms} J.,  {Dove} J.~B.,   {Begelman}
  M.~C.,  1999, \mn@doi [\apj] {10.1086/306610}, \href
  {https://ui.adsabs.harvard.edu/abs/1999ApJ...510..874N} {510, 874}

\bibitem[\protect\citeauthoryear{{Poutanen}, {Veledina}  \&
  {Zdziarski}}{{Poutanen} et~al.}{2018}]{2018A&A...614A..79P}
{Poutanen} J.,  {Veledina} A.,   {Zdziarski} A.~A.,  2018, \mn@doi [\aap]
  {10.1051/0004-6361/201732345}, \href
  {https://ui.adsabs.harvard.edu/abs/2018A&A...614A..79P} {614, A79}

\bibitem[\protect\citeauthoryear{{Rawat} \& {et al.,}}{{Rawat} \& {et
  al.,}}{2022}]{rawat}
{Rawat} D.,  {et al.,} 2022, submitted to MNRAS

\bibitem[\protect\citeauthoryear{{Reig}, {Belloni}, {van der Klis},
  {M{\'e}ndez}, {Kylafis}  \& {Ford}}{{Reig}
  et~al.}{2000}]{2000ApJ...541..883R}
{Reig} P.,  {Belloni} T.,  {van der Klis} M.,  {M{\'e}ndez} M.,  {Kylafis}
  N.~D.,   {Ford} E.~C.,  2000, \mn@doi [\apj] {10.1086/309469}, \href
  {https://ui.adsabs.harvard.edu/abs/2000ApJ...541..883R} {541, 883}

\bibitem[\protect\citeauthoryear{{Remillard} \& {McClintock}}{{Remillard} \&
  {McClintock}}{2006}]{2006ARA&A..44...49R}
{Remillard} R.~A.,  {McClintock} J.~E.,  2006, \mn@doi [\araa]
  {10.1146/annurev.astro.44.051905.092532}, \href
  {https://ui.adsabs.harvard.edu/abs/2006ARA&A..44...49R} {44, 49}

\bibitem[\protect\citeauthoryear{{Remillard}, {Morgan}, {McClintock}, {Bailyn}
  \& {Orosz}}{{Remillard} et~al.}{1999}]{1999ApJ...522..397R}
{Remillard} R.~A.,  {Morgan} E.~H.,  {McClintock} J.~E.,  {Bailyn} C.~D.,
  {Orosz} J.~A.,  1999, \mn@doi [\apj] {10.1086/307606}, \href
  {https://ui.adsabs.harvard.edu/abs/1999ApJ...522..397R} {522, 397}

\bibitem[\protect\citeauthoryear{{Russell}, {Miller-Jones}, {Maccarone},
  {Yang}, {Fender}  \& {Lewis}}{{Russell} et~al.}{2011}]{2011ApJ...739L..19R}
{Russell} D.~M.,  {Miller-Jones} J.~C.~A.,  {Maccarone} T.~J.,  {Yang} Y.~J.,
  {Fender} R.~P.,   {Lewis} F.,  2011, \mn@doi [\apjl]
  {10.1088/2041-8205/739/1/L19}, \href
  {https://ui.adsabs.harvard.edu/abs/2011ApJ...739L..19R} {739, L19}

\bibitem[\protect\citeauthoryear{{Russell}, {Miller-Jones}, {Sivakoff},
  {Tetarenko}  \& {Jacpot Xrb Collaboration}}{{Russell}
  et~al.}{2017}]{2017ATel10711....1R}
{Russell} T.~D.,  {Miller-Jones} J.~C.~A.,  {Sivakoff} G.~R.,  {Tetarenko}
  A.~J.,   {Jacpot Xrb Collaboration} 2017, The Astronomer's Telegram, \href
  {https://ui.adsabs.harvard.edu/abs/2017ATel10711....1R} {10711, 1}

\bibitem[\protect\citeauthoryear{{Russell} et~al.,}{{Russell}
  et~al.}{2019}]{2019ApJ...883..198R}
{Russell} T.~D.,  et~al., 2019, \mn@doi [\apj] {10.3847/1538-4357/ab3d36},
  \href {https://ui.adsabs.harvard.edu/abs/2019ApJ...883..198R} {883, 198}

\bibitem[\protect\citeauthoryear{{Russell} et~al.,}{{Russell}
  et~al.}{2020}]{2020MNRAS.498.5772R}
{Russell} T.~D.,  et~al., 2020, \mn@doi [\mnras] {10.1093/mnras/staa2650},
  \href {https://ui.adsabs.harvard.edu/abs/2020MNRAS.498.5772R} {498, 5772}

\bibitem[\protect\citeauthoryear{{Sridhar}, {Bhattacharyya}, {Chandra}  \&
  {Antia}}{{Sridhar} et~al.}{2019}]{2019MNRAS.487.4221S}
{Sridhar} N.,  {Bhattacharyya} S.,  {Chandra} S.,   {Antia} H.~M.,  2019,
  \mn@doi [\mnras] {10.1093/mnras/stz1476}, \href
  {https://ui.adsabs.harvard.edu/abs/2019MNRAS.487.4221S} {487, 4221}

\bibitem[\protect\citeauthoryear{{Stella} \& {Vietri}}{{Stella} \&
  {Vietri}}{1998}]{1998ApJ...492L..59S}
{Stella} L.,  {Vietri} M.,  1998, \mn@doi [\apjl] {10.1086/311075}, \href
  {https://ui.adsabs.harvard.edu/abs/1998ApJ...492L..59S} {492, L59}

\bibitem[\protect\citeauthoryear{{Stiele} \& {Kong}}{{Stiele} \&
  {Kong}}{2018}]{2018ApJ...868...71S}
{Stiele} H.,  {Kong} A.~K.~H.,  2018, \mn@doi [\apj]
  {10.3847/1538-4357/aae7d3}, \href
  {https://ui.adsabs.harvard.edu/abs/2018ApJ...868...71S} {868, 71}

\bibitem[\protect\citeauthoryear{{Stirling}, {Spencer}, {de la Force},
  {Garrett}, {Fender}  \& {Ogley}}{{Stirling}
  et~al.}{2001}]{2001MNRAS.327.1273S}
{Stirling} A.~M.,  {Spencer} R.~E.,  {de la Force} C.~J.,  {Garrett} M.~A.,
  {Fender} R.~P.,   {Ogley} R.~N.,  2001, \mn@doi [\mnras]
  {10.1046/j.1365-8711.2001.04821.x}, \href
  {https://ui.adsabs.harvard.edu/abs/2001MNRAS.327.1273S} {327, 1273}

\bibitem[\protect\citeauthoryear{{Svoboda}, {Dov{\v{c}}iak}, {Goosmann},
  {Jethwa}, {Karas}, {Miniutti}  \& {Guainazzi}}{{Svoboda}
  et~al.}{2012}]{2012A&A...545A.106S}
{Svoboda} J.,  {Dov{\v{c}}iak} M.,  {Goosmann} R.~W.,  {Jethwa} P.,  {Karas}
  V.,  {Miniutti} G.,   {Guainazzi} M.,  2012, \mn@doi [\aap]
  {10.1051/0004-6361/201219701}, \href
  {https://ui.adsabs.harvard.edu/abs/2012A&A...545A.106S} {545, A106}

\bibitem[\protect\citeauthoryear{{Tagger} \& {Pellat}}{{Tagger} \&
  {Pellat}}{1999}]{1999A&A...349.1003T}
{Tagger} M.,  {Pellat} R.,  1999, \aap, \href
  {https://ui.adsabs.harvard.edu/abs/1999A&A...349.1003T} {349, 1003}

\bibitem[\protect\citeauthoryear{{Tao} et~al.,}{{Tao}
  et~al.}{2018}]{2018MNRAS.480.4443T}
{Tao} L.,  et~al., 2018, \mn@doi [\mnras] {10.1093/mnras/sty2157}, \href
  {https://ui.adsabs.harvard.edu/abs/2018MNRAS.480.4443T} {480, 4443}

\bibitem[\protect\citeauthoryear{{Tingay} et~al.,}{{Tingay}
  et~al.}{1995}]{1995Natur.374..141T}
{Tingay} S.~J.,  et~al., 1995, \mn@doi [\nat] {10.1038/374141a0}, \href
  {https://ui.adsabs.harvard.edu/abs/1995Natur.374..141T} {374, 141}

\bibitem[\protect\citeauthoryear{{Van der Klis}}{{Van der
  Klis}}{1989}]{1989ASIC..262...27V}
{Van der Klis} M.,  1989, in {{\"O}gelman} H.,  {van den Heuvel} E.~P.~J.,
  eds,  NATO Advanced Study Institute (ASI) Series C Vol. 262, Timing Neutron
  Stars. p.~27

\bibitem[\protect\citeauthoryear{{Vaughan} \& {Nowak}}{{Vaughan} \&
  {Nowak}}{1997}]{1997ApJ...474L..43V}
{Vaughan} B.~A.,  {Nowak} M.~A.,  1997, \mn@doi [\apjl] {10.1086/310430}, \href
  {https://ui.adsabs.harvard.edu/abs/1997ApJ...474L..43V} {474, L43}

\bibitem[\protect\citeauthoryear{{Verner}, {Ferland}, {Korista}  \&
  {Yakovlev}}{{Verner} et~al.}{1996}]{1996ApJ...465..487V}
{Verner} D.~A.,  {Ferland} G.~J.,  {Korista} K.~T.,   {Yakovlev} D.~G.,  1996,
  \mn@doi [\apj] {10.1086/177435}, \href
  {https://ui.adsabs.harvard.edu/abs/1996ApJ...465..487V} {465, 487}

\bibitem[\protect\citeauthoryear{{Wang} et~al.,}{{Wang}
  et~al.}{2021}]{2021ApJ...910L...3W}
{Wang} J.,  et~al., 2021, \mn@doi [\apjl] {10.3847/2041-8213/abec79}, \href
  {https://ui.adsabs.harvard.edu/abs/2021ApJ...910L...3W} {910, L3}

\bibitem[\protect\citeauthoryear{{Wilkins} \& {Fabian}}{{Wilkins} \&
  {Fabian}}{2012}]{2012MNRAS.424.1284W}
{Wilkins} D.~R.,  {Fabian} A.~C.,  2012, \mn@doi [\mnras]
  {10.1111/j.1365-2966.2012.21308.x}, \href
  {https://ui.adsabs.harvard.edu/abs/2012MNRAS.424.1284W} {424, 1284}

\bibitem[\protect\citeauthoryear{{Wilms}, {Allen}  \& {McCray}}{{Wilms}
  et~al.}{2000}]{2000ApJ...542..914W}
{Wilms} J.,  {Allen} A.,   {McCray} R.,  2000, \mn@doi [\apj] {10.1086/317016},
  \href {https://ui.adsabs.harvard.edu/abs/2000ApJ...542..914W} {542, 914}

\bibitem[\protect\citeauthoryear{{Xu} et~al.,}{{Xu}
  et~al.}{2018}]{2018ApJ...852L..34X}
{Xu} Y.,  et~al., 2018, \mn@doi [\apjl] {10.3847/2041-8213/aaa4b2}, \href
  {https://ui.adsabs.harvard.edu/abs/2018ApJ...852L..34X} {852, L34}

\bibitem[\protect\citeauthoryear{{You}, {Bursa}  \& {{\.Z}ycki}}{{You}
  et~al.}{2018}]{2018ApJ...858...82Y}
{You} B.,  {Bursa} M.,   {{\.Z}ycki} P.~T.,  2018, \mn@doi [\apj]
  {10.3847/1538-4357/aabd33}, \href
  {https://ui.adsabs.harvard.edu/abs/2018ApJ...858...82Y} {858, 82}

\bibitem[\protect\citeauthoryear{{You} et~al.,}{{You}
  et~al.}{2021}]{2021NatCo..12.1025Y}
{You} B.,  et~al., 2021, \mn@doi [Nature Communications]
  {10.1038/s41467-021-21169-5}, \href
  {https://ui.adsabs.harvard.edu/abs/2021NatCo..12.1025Y} {12, 1025}

\bibitem[\protect\citeauthoryear{{Zdziarski}, {Johnson}  \&
  {Magdziarz}}{{Zdziarski} et~al.}{1996}]{1996MNRAS.283..193Z}
{Zdziarski} A.~A.,  {Johnson} W.~N.,   {Magdziarz} P.,  1996, \mn@doi [\mnras]
  {10.1093/mnras/283.1.193}, \href
  {https://ui.adsabs.harvard.edu/abs/1996MNRAS.283..193Z} {283, 193}

\bibitem[\protect\citeauthoryear{{Zhang}, {Lu}, {Zhang}  \& {Li}}{{Zhang}
  et~al.}{2014}]{2014SPIE.9144E..21Z}
{Zhang} S.,  {Lu} F.~J.,  {Zhang} S.~N.,   {Li} T.~P.,  2014, in {Takahashi}
  T.,  {den Herder} J.-W.~A.,   {Bautz} M.,  eds,  Society of Photo-Optical
  Instrumentation Engineers (SPIE) Conference Series Vol. 9144, Space
  Telescopes and Instrumentation 2014: Ultraviolet to Gamma Ray. p. 914421,
  \mn@doi{10.1117/12.2054144}

\bibitem[\protect\citeauthoryear{{Zhang} et~al.,}{{Zhang}
  et~al.}{2020a}]{2020MNRAS.494.1375Z}
{Zhang} L.,  et~al., 2020a, \mn@doi [\mnras] {10.1093/mnras/staa797}, \href
  {https://ui.adsabs.harvard.edu/abs/2020MNRAS.494.1375Z} {494, 1375}

\bibitem[\protect\citeauthoryear{{Zhang} et~al.,}{{Zhang}
  et~al.}{2020b}]{2020MNRAS.499..851Z}
{Zhang} L.,  et~al., 2020b, \mn@doi [\mnras] {10.1093/mnras/staa2842}, \href
  {https://ui.adsabs.harvard.edu/abs/2020MNRAS.499..851Z} {499, 851}

\bibitem[\protect\citeauthoryear{{{\.Z}ycki}, {Done}  \& {Smith}}{{{\.Z}ycki}
  et~al.}{1999}]{1999MNRAS.309..561Z}
{{\.Z}ycki} P.~T.,  {Done} C.,   {Smith} D.~A.,  1999, \mn@doi [\mnras]
  {10.1046/j.1365-8711.1999.02885.x}, \href
  {https://ui.adsabs.harvard.edu/abs/1999MNRAS.309..561Z} {309, 561}

\makeatother
\end{thebibliography}



\appendix




\bsp	
\label{lastpage}
\end{document}